\documentclass{aa}
\usepackage{graphicx}
\usepackage{txfonts}
\usepackage{float,psfig}
\usepackage{aalongtable}
\usepackage{lscape}

\def\chandra{\it Chandra\ \rm}
\def\xmm{\it XMM-Newton\ \rm}
\def\sax{\it BeppoSAX\ \rm}

\begin{document}

\title{Cross-calibrating X-ray detectors with clusters of galaxies: an IACHEC study} 

\author{J. Nevalainen\inst{1,2}
          \and
          L. David\inst{3}
          \and
          M. Guainazzi\inst{4}
         }

   \offprints{J. Nevalainen}

   \institute{Department of Physics, Dynamicum, P.O.Box 48, 00014 University of Helsinki, Finland
         \and
              Finnish Centre for  Astronomy with ESO, University of Turku, V\"ais\"al\"antie 20, FI-21500 Piikki\"o, Finland
         \and
             Harvard-Smithsonian Center for Astrophysics, 60 Garden Street, Cambridge, USA
          \and
          European Space Astronomy Center of ESA, P.O.Box 78, Villanueva de la Ca\~nada, E-28691 Madrid, Spain 
             }

   \date{Received; accepted}

\abstract
{}
{We used a sample of 11 nearby relaxed clusters of galaxies observed with the X-ray instruments \xmm (EPIC) pn and 
MOS, \chandra ACIS-S and ACIS-I and \sax MECS to examine the cross-calibration of the energy dependence and
normalisation of the effective area of these instruments as of December 2009. 
We also examined the Fe XXV/XXVI line ratio temperature measurement method for the pn and MOS.}
{We performed X-ray spectral analysis on the \xmm and \chandra data for a sample of 11 clusters. We obtained the
information for \sax from DeGrandi \& Molendi (2002). We compared the spectroscopic results obtained with different 
instruments for the same clusters in order to examine possible systematic calibration effects between the instruments.}
{We did not detect any significant systematic differences between the temperatures derived in the 2--7 keV band using the 
different instruments. Also, the EPIC temperatures derived from the bremsstrahlung continuum agreed with those obtained 
from the Fe XXV/XXVI emission line ratio, implying that the energy dependence of the hard band effective area of the 
above instruments is accurately calibrated. This also indicates that deviations from ionisation equilibrium
and a Maxwellian electron velocity distribution are negligible in the regions studied in the cluster sample. We thus 
consider the IACHEC sample of clusters of galaxies as standard candles for the calibration of the energy dependence of 
the hard band (2-7 keV) effective area of X-ray telescopes. On the other hand, the hard band EPIC/ACIS fluxes disagreed by 
5--10\% (i.e. at 6--25$\sigma$ level) which 
indicates a similar level of uncertainty in the normalisations of the effective areas of these instruments in the 2--7 keV band.
In the soft energy band (0.5--2.0 keV) there are greater cross-calibration differences between EPIC and ACIS.
We found an energy-dependent increase of ACIS versus pn bias in the cross-calibration of the effective area 
by $\sim$ 10\% in the 0.5 -- 2.0~keV band. This amounts to a 
systematic difference of $\sim$20\% in the temperatures measured by the ACIS and the EPIC-pn cameras in this band.
Due to the high statistical weight of the soft band data, the 0.5--7.0 keV band 
temperature measurements of clusters of galaxies with EPIC/\xmm or ACIS/\chandra are uncertain by $\sim$10--15\% on 
average. These uncertainties will also affect the analysis of the wide band continuum spectra of other types of objects 
using ACIS or EPIC.
} 
{}


\authorrunning{J. Nevalainen et al.}
\titlerunning{IACHEC cluster calibration}

\maketitle

\section{Introduction}
This paper belongs to a series of studies aimed at assessing the cross-calibration status among operational and 
past X-ray missions promoted by the {\it International Astronomical Consortium for High Energy Calibration 
(IACHEC)\footnote{http://web.mit.edu/iachec}}.  
In this work, we mainly explore the use of clusters of galaxies for the cross-calibration of the effective area 
(defined here as the product of the mirror effective area, the detector quantum efficiency and the filter transmission). 
Unlike many astronomical targets used for X-ray calibration, clusters of galaxies are stable. 
There is no simultaneity requirement for observing clusters of galaxies with different instruments.
Thus, it is possible to obtain large cross-calibration samples of clusters of galaxies which enables a statistically
meaningful analysis of systematic effects between X-ray instruments.
Clusters of galaxies are relatively bright (L$\sim 10^{45}$ erg s$^{-1}$) and thus, the measurements can be performed
to a high statistical precision which translates into equally precise information on the calibration.
The brightness is not high enough to introduce pile-up problems which is the case with many types of calibration 
sources. On the other hand, the extended nature of clusters of galaxies introduces additional complexity due to e.g. 
point spread function and vignetting.

The X-ray spectra of galaxy clusters can typically be described by the combination of collisionally ionised plasma 
emission components, whose physics is fairly well known. While the energy dependence (i.e. the shape) of 
the effective area primarily impacts the determination of the cluster temperature, the normalisation of the 
effective area affects the determination of the emission measure and flux. Comparing temperatures and fluxes measured
by different instruments for the same cluster yields information on the cross-calibration of the effective areas.
For the hottest clusters, the H-like and He-like FeK$\alpha$ lines (Fe XXV and Fe XXVI) are measurable and thus the 
temperature-dependent Fe XXV/XXVI line flux ratio provides an additional diagnostic tool for the cross-calibration. 
These lines cover a narrow energy band where the uncertainties of the calibration of the shape of the effective area 
do not substantially affect the measurement.

\begin{table*}
\caption{\label{clustinfo2_tab}Information on the observations of the sample.}
\centering
{\small
\begin{tabular}{l|ccccccc|cccc}
\hline\hline
        &  \multicolumn{7}{c}{\xmm}              & \multicolumn{4}{c}{\chandra}                                \\
name    &  ID   & Obs. date     & Optical filter & \multicolumn{3}{c}{exposure time (ks)} & lc filter\tablefootmark{a} &  ID & Obs. date  & \multicolumn{2}{c}{exposure time (ks)}                  \\  
        &            & yyyy-mm-dd & & pn  & MOS1 & MOS2 &           &        & yyyy-mm-dd & ACIS-S & ACIS-I          \\ 
\hline
A1795   & 0097820101 & 2000-06-26 & thin   & 23  & 37   & 34   & flare     & 6160   & 2005-03-20 & 14     &  --    \\
A2029   & 0111270201 & 2002-08-25 & thin   & 8   & 12   & 12   & flare     & 4977   & 2004-01-09 & 78     &  --    \\
A2052   & 0109920101 & 2000-08-21 & thin   & 23  & 30   & 30   & $\pm$20\% & 5807   & 2006-03-26 & 127    &  --    \\
A2199   & 0008030201 & 2002-07-04 & thin   & 12  & 14   & 14   & $\pm$20\% & 497    & 2000-05-13 & 16     &  --    \\
A262    & 0109980101 & 2001-01-16 & thin   & 16  & 23   & 24   & $\pm$20\% & 2215   & 2001-08-03 & 30     &  --    \\
A3112   & 0105660101 & 2000-12-24 & medium & 17  & 22   & 23   & $\pm$20\% & 2516   & 2001-09-15 & 11     &  --    \\
A3571   & 0086950201 & 2002-07-29 & medium & 7   & 19   & 20   & flare     & 4203   & 2003-07-31 & 8      &  --    \\
A85     & 0065140101 & 2002-01-07 & medium & 9   & 12   & 12   & $\pm$20\% & 904    & 2000-08-19 & --     &  38    \\
Coma    & 0153750101 & 2001-12-04 & medium & 17  & 21   & 21   & $\pm$20\% & 9714   & 2008-03-20 & --     &  30    \\
HydraA  & 0109980301 & 2000-12-08 & thin   & 7   & 18   & 19   & flare     & 4969   & 2004-01-13 & 88     & --     \\
MKW3S   & 0109930101 & 2000-08-22 & thin   & 28  & 35   & 35   & $\pm$20\% & 900    & 2000-04-03 & --     & 56     \\ 
\hline
\end{tabular}}
\tablefoot{
\tablefoottext{a}{The method for cleaning the particle flare periods: ``$\pm$20\%'' refers to $\pm$20\% filter around the 
quiescence while ``flare'' filter denotes clusters where the mild flares are allowed.}
}
\end{table*}

In this work, we use the clusters A1795, A2029, A2052, A2199, A262, A3112, A3571, A85, Coma, HydraA and MKW3S observed 
with satellites (instruments) \xmm (pn and MOS) , \chandra (ACIS-S and ACIS-I) and \sax 
(MECS) (see Table \ref{clustinfo2_tab}). 
We compared the temperatures and fluxes in order to determine the status of the cross-calibration of these instruments 
for the implementation of the calibration information as of December 2009.

\section{The cluster properties and selected regions}
\label{regions}
The chosen clusters are nearby (z$<$0.08), and bright  (the flux in the 2--7 keV band $\sim 10^{-12} - 10^{-11}$ erg 
s$^{-1}$ cm$^{-2}$) and most host a cool core (see Table \ref{clustinfo_tab}). These clusters exhibit no major merger 
signatures in the central regions. The Galactic hydrogen column density is less than $6 \times 10^{20}$~cm$^{-2}$
in all of the clusters. The temperature of the gas ranges from $\sim$2 keV to $\sim$10 keV in the cluster sample.

The most significant temperature variation in the relaxed cool core clusters is due to the radial decline of the  
gas temperature and the central cooling. To minimise the effect of the radial dependence of the gas temperature
between observations by different instruments, we aim to extract spectra within the same annular regions near the 
peak of the cluster temperature profile for a given cluster.
This is complicated because of  the different obscured areas due to, e.g., CCD gaps in the different instruments. 
The full annulus set by the chosen inner and outer extraction radii is obscured by $\sim$15\% (pn) and  $\sim$5\% (MOS), 
on average, in our sample. ACIS-I observations of A85 and Coma cover $\sim$85\% of the full annulus, while the ACIS-I 
observation of MKW3S and all the ACIS-S observations cover 100\% of the full annulus. We assume in this work that the 
differences in the covered areas do not produce significant effects on the derived gas temperatures. 
We minimise the emission from bright point sources by excluding circular regions with radii of 30$''$ centered on 
the same sky positions in EPIC and ACIS (see Table \ref{pointsrc.tab}).

We treated the \xmm EPIC instruments pn, MOS1 and MOS2 separately, except in some relevant cases where we analysed the 
combined MOS1 and MOS2 data for a given cluster and thus formed a single comparison group we identify as MOS. When comparing 
the temperatures from EPIC instruments only, we used the data extracted from an annular region with an outer radius, 
r$_{\rm out}$, of 6 arcmin\footnote{The size of the inner extraction radius depends on several issues (e.g. PSF scatter
and central cooling) and will be discussed in detail in Sects. \ref{centralcool} and \ref{psf}}.

\begin{table}
\caption{\label{clustinfo_tab}General information on the sample.}
\centering
\begin{tabular}{lcccc}
\hline\hline
        &                      &        &             &            \\
name    & NH\tablefootmark{a}             & z\tablefootmark{b} & R.A.(J2000)\tablefootmark{c} & Dec (J2000)\tablefootmark{c}         \\
        & [$10^{20}$ cm$^{-2}$] &        & hh mm ss    & $^{\circ}$ \ ' \ ''\   \\
\hline
A1795   & 1.19               & 0.0625 & 13 48 53.0  & 26 35 25.0   \\
A2029   & 3.25               & 0.0773 & 15 10 56.2  & 05 44 40.7   \\
A2052   & 2.71               & 0.0355 & 15 16 44.5  & 07 01 19.7   \\
A2199   & 0.89               & 0.0302 & 16 28 38.5  & 39 33 06.8   \\
A262    & 5.67               & 0.0163 & 01 52 46.0  & 36 09 09.1   \\
A3112   & 1.33               & 0.0753 & 03 17 57.7  & -44 14 18.3  \\
A3571   & 4.25               & 0.0391 & 13 47 28.6  & -32 51 54.8  \\
A85     & 2.78               & 0.0551 & 00 41 50.5  & -09 18 10.8  \\
Coma    & 0.87               & 0.0231 & 12 59 35.7  &  27 57 34.0  \\
HydraA  & 4.60               & 0.0539 & 09 18 05.7  &  -12 05 43.5 \\
MKW3S   & 2.68               & 0.0450 & 15 21 51.4  & 07 42 22.9   \\
\hline
\end{tabular}
\tablefoot{
\tablefoottext{a}{LAB weighted average (Kalberla et al., 2005).} 
\tablefoottext{b}{From NASA Extragalactic Database.}
\tablefoottext{c}{The adopted coordinates of the cluster centre based on the location of the X-ray brightness peak 
in the \xmm pn images}.
}
\end{table}

While the full sample of 11 clusters is observed simultaneously with the \xmm EPIC instruments, only a few clusters 
have been observed by both ACIS-S and ACIS-I. Our analysis of the ACIS-S and ACIS-I data of Coma and A1795 showed that
the broad band temperatures are consistent within the 3\% statistical uncertainties and that the broad band fluxes may 
differ by up to 2-3\%. Due to these close agreements, we combine the ACIS-S and ACIS-I results as a single 
comparison group we identify as ACIS. This group contains all of the 11 clusters. The small FOV of ACIS-S 
($\sim 8'\times8'$) sets the limit for the outer radii r$_{\rm out}$ of the spectrum extraction. Depending on the 
placement of the cluster centre in the chip, 
the outer radii vary between 2--3$'$. Thus, for EPIC/ACIS-S comparison we used data for a given cluster 
extracted from an annular region with the outer radius  set by ACIS-S (see Table \ref{clustinfo3.tab}).

Furthermore, for comparison with \sax MECS, we will use the published 2--10 keV band best-fit single temperature MEKAL 
values obtained in the central 0--2--4--6 arcmin regions (deGrandi \& Molendi, 2002). We refer to this paper for
the details of the data analysis, including the correction for the point spread function. For comparison with MECS, 
we extracted pn spectra from the same regions as used for MECS. Due to the limitations of the exposure time and the 
background, we were able to use 12 regions in 6 clusters in common with the pn and MECS (see Sect. \ref{xmm-bepposax}).

\begin{table*}
\caption[]{Point source coordinates}
\label{pointsrc.tab}
{
\begin{tabular}{llll}
\hline\hline
  &   & &   \\
cluster &   \multicolumn{3}{c}{R.A.(J2000) ; Dec (J2000)} \\ 
         & \multicolumn{3}{c}{hh mm ss   ; $^{\circ}$ \ ' \ ''} \\ 
\hline
A1795    & 13 48 35.1 ;  26 31 08.5 &                          &                           \\
A2029    & 15 11 06.3 ;  05 41 21.7 & 15 10 37.3 ; 05 48 13.6  & 15 11 00.6 ;  05 49 22.1  \\
A2052    & 15 16 44.6 ;  07 05 13.8 & 15 16 32.0 ; 06 58 51.1  & 15 16 55.0 ;  07 01 56.2  \\
A2199    & 16 28 26.2 ;  39 33 52.6 & 16 28 24.0 ; 39 33 21.1  & 16 29 07.0 ;  39 32 39.3  \\
A262     & 01 52 39.4 ;  36 07 24.2 & 01 52 39.6 ; 36 10 16.5  & 01 52 22.3 ;  36 06 43.0  \\
         & 01 52 32.0 ;  36 05 14.3 & 01 53 07.1 ; 36 09 01.2  & 01 53 07.5 ;  36 12 21.5  \\
         & 01 52 39.3 ;  36 12 32.3 & 01 52 21.6 ; 36 09 04.6  & 01 53 02.9 ;  36 12 41.0  \\
A3112    & 03 18 04.3 ; -44 13 48.3 & 03 18 02.4 ; -44 16 43.3 & 03 17 23.6 ; -44 15 06.7  \\
         & 03 17 32.3 ; -44 17 34.5 & 03 17 52.0 ; -44 18 29.1 & 03 18 02.1 ; -44 18 01.8  \\
A3571    & 13 47 33.0 ; -32 52 59.8 & 13 47 29.0 ; -32 48 34.9 & 13 47 18.7 ; -32 49 10.1  \\
A85      & 00 41 40.5 ; -09 19 47.9 & 00 41 47.9 ; -09 20 44.0 & 00 42 00.3 ; -09 19 30.2  \\
         & 00 41 30.3 ; -09 15 47.1 &                          &                           \\
Coma     & 12 59 16.6 ;  27 53 42.4 &                          &                           \\
HydraA   & 09 17 58.4 ; -12 04 49.6 & 09 18 05.4 ; -12 08 00.5 & 09 18 15.4 ; -12 05 27.0  \\ 
         & 09 17 53.5 ; -12 10 37.0 & 09 17 50.0 ; -12 04 29.8 & 09 18 22.4 ; -12 04 35.3  \\
         & 09 18 23.5 ; -12 02 22.2 & 09 18 09.1 ; -12 00 58.5 & 09 18 19.1 ; -12 02 53.0  \\
MKW3S    & 15 21 56.2 ;  07 37 09.4 & 15 21 39.7 ;  07 38 33.0 & 15 21 45.7 ;  07 39 34.9  \\ 
         & 15 22 12.0 ;  07 40 01.8 & 15 22 07.9 ;  07 45 22.5 & 15 21 56.1 ;  07 37 03.0  \\
         & 15 21 52.3 ;  07 36 33.4 & 15 21 36.3 ;  07 40 37.5 & 15 21 54.1 ;  07 47 03.2  \\ 
\hline
\end{tabular}}
\end{table*}

\section{Data processing}
For each cluster we used the data obtained by such on-axis pointing (see Table \ref{clustinfo2_tab}) which was publicly 
available in May 2007 and has the largest useful exposure time after the flare-filtering  (except that for Coma we used 
an ACIS-I observation from the year 2008).

\subsection{\xmm}
We processed the raw \xmm data with the SASv9.0 tools epchain and emchain with the default parameters in order to 
produce the event files. We used the latest calibration information as of December 2009. We also generated the simulated 
out-of-time event file, which we later used to subtract the events from the pn spectra registered during the readout of a
CCD. We filtered the event files excluding bad pixels and CCD gaps. We further filtered the event files including only 
patterns 0--4 (pn) and 0--12 (MOS). We used the evselect-3.60.3 tool to extract spectra, images, and light curves, while
excluding the regions contaminated by bright point sources. We used the rmfgen-1.55.1 and arfgen-1.76.4 tools to produce
the energy redistribution files and the effective area files. When running the arfgen tool, we used an extended source 
configuration and supplied an \xmm image of the cluster in detector coordinates, binned in 0.5 arcmin pixels,
for weighting the response.

\subsection{\chandra}
The data for each cluster were extracted from the \chandra archive and reprocessed with the CIAO 4.2 version of 
$acis\_process\_events$ tool. To examine the effects of recent calibration updates, we analysed the data using 
CALDB 3.4, CALDB 4.1.1 (which included an updated HRMA effective area) and CALDB 4.2.0 (which includes an updated model for
the contamination build-up on the ACIS filters). To filter out background flares, we generated light curves in the 
2.5-7.0~keV and 9.0-12.0~keV energy bands using the data from the S1 chip for ACIS-S3 observations and from the S2 chip for 
ACIS-I  observations. All times with background rates exceeding 20\% of the mean were excluded from further analysis.

\section{Background spectra}
\label{background}
Our sample consists of nearby clusters that fill the whole FOV of the detectors and thus a local estimate for the 
background spectrum is not possible.

\subsection{\xmm EPIC}
For the \xmm EPIC analysis, we used blank sky--based estimates for the total sky+particle background spectra from 
Nevalainen et al. (2005). Utilising the $>$10 keV ($>$ 9.5 keV) light curves for the pn (MOS) we 
only accepted data from such periods when the count rate is within $\pm$20\% of the quiescent level. However, 
in some cases, this led to very low exposure times.  In order to improve the statistical precision for these clusters, we 
accepted the data accumulated during periods of mild flares, i.e. when the $>$10 keV count rate level was lower than 2.0 
$\times$ the quiescent rate (see Table \ref{clustinfo2_tab}). 

In order to account for the variability of the instrumental background due to cosmic rays, we used the sample of EPIC 
exposures taken with a CLOSED filter (Nevalainen et al. 2005) to extract particle background spectra at the same detector 
regions as used for the cluster data. We included this additional component in the fits after adjusting its normalisation
so that the total background count rate prediction in the 10--14 keV (9.5--12 keV) band for the pn (MOS) matches that in the
cluster observation\footnote{Due to the negligible effective area the cluster emission is insignificant at these energies}. 

\begin{table*}
\caption{\label{clustinfo3.tab}Properties of the clusters.}
\centering
\begin{tabular}{lcccccccccc}
\hline\hline
          &                       &                     &              &             &              &             &           &                &          \\
name      & r$_\mathrm{in}$\tablefootmark{a}  & r$_\mathrm{out}$\tablefootmark{a}  & I$_{0,1}$/I$_{0,2}$\tablefootmark{b}  & r$_\mathrm{core,1}$\tablefootmark{b}  & $\beta_{1}$\tablefootmark{b} & r$_\mathrm{core,2}$\tablefootmark{b}  & $\beta_{2}$\tablefootmark{b} & r$_{500}$\tablefootmark{c} & 
f$_\mathrm{non-isot}$\tablefootmark{d} & f$_\mathrm{psf}$\tablefootmark{e}  \\
          & $'$       & $'$       &                     & $''$         &             & $''$         &             &  $'$      & \%             & \%       \\
\hline
A1795     & 1.5  & 2.7  & 2.7        & 13.7  & 0.41 & 49.1     & 0.67      & 13  & 7  & 0.5      \\
A2029     & 1.5  & 2.5  & 3.8        & 3.0   & 0.42 & 33.8     & 0.53      & 11  & 9  & 0.3      \\
A2052     & 1.7  & 2.5  & 8.5        & 42.1  & 0.97 & 150.0    & 0.65      & 23  & 5  & 6        \\
A2199     & 2.0  & 2.9  & 6.5        & 3.3   & 0.43 & 54.2     & 0.47      & 27  & 6  & 6        \\
A262      & 1.6  & 2.7  & 8.4        & 20.0  & 0.68 & 100.0    & 0.44      & 49  & 4  & 5        \\
A3112     & 1.5  & 2.9  & 8.3        & 4.7   & 0.46 & 28.8     & 0.50      & 11  & 17 & 0.3      \\
A3571     & 0.0  & 2.1  & 4.7        & 11.3  & 0.27 & 122.3    & 0.67      & 28  & 5  & $\ldots$ \\
A85       & 1.5  & 3.0  & 13.3       & 12.1  & 0.51 & 137.0    & 0.7       & 15  & 6  & 0.4      \\
Coma      & 1.0  & 5.0  & $\ldots$   & 630   & 0.75 & $\ldots$ & $\ldots$  & 50  & 8  & $\ldots$ \\
HydraA    & 1.5  & 2.7  & $\ldots$   & 17.6  & 0.49 & $\ldots$ & $\ldots$  & 15  & 9  & 0.9      \\
MKW3S     & 1.5  & 2.5  & $\ldots$   & 25.7  & 0.44 & $\ldots$ & $\ldots$  & 18  & 10 & 2        \\
\hline
\end{tabular}
\tablefoot{
\tablefoottext{a}{The adopted inner and outer radii for the annular spectrum extraction region for EPIC--ACIS comparison.}
\tablefoottext{b}{The $\beta$-profile parameters are obtained in this work, except that for Coma we adopted the values 
from Briel et al. (1992)}
\tablefoottext{c}{Estimated using the formula in Vikhlinin et al. (2006).}
\tablefoottext{d}{The contribution from outside the isothermal region (r $> 0.3$ r$_{500}$) to 
the projected extraction region.}
\tablefoottext{e}{The fraction of the emission within r$_{\rm in}$ and r$_{\rm out}$ due to the PSF scatter from the central cool 
region in pn.}
}
\end{table*}

We used the blank sky study of Nevalainen et al. (2005) to estimate the relative uncertainty of our adopted background 
model in different energy bands and instruments (varying between 5\% in the hard band for MOS and 20\% in the soft band 
for the pn).
We varied the background model by the appropriate amount and interpreted the change in the best-fit temperature as a 
systematic effect due to the background uncertainties. We propagated this uncertainty by adding it in quadrature to the 
statistical uncertainties of the temperature. The background uncertainty has some effect on the temperature measurement 
in the hard band (at $\sim$ 2\% level) for the faintest clusters (A262, A2052, and MKW3S), where the background is the 
highest (10--15\% of the cluster signal). In the soft band, the background level is only a few per cent of that of the 
cluster signal, and thus the effect of the background uncertainty is negligible.  

In this work, we assumed that the sky background emission is absorbed by the same amount of neutral hydrogen as 
in the cluster field. Thus, in principle, the difference between the hydrogen column density measured along the 
line-of-sight to the cluster and the average column density affecting the blank sky field pointings may bias the  
temperature measurements. We used A2052 as a conservative example of the column density bias, because it is the faintest
object in our sample in the soft band (the background flux is $\sim$6\% of the cluster flux in the soft band). Assuming 
conservatively that the column density of the blank sky sample has the minimum value of all the fields
used to generate the background file, i.e. 
$0.6 \times 10^{20}$~cm$^{-2}$ (Nevalainen et al., 2005), the fraction of the absorbed background flux differs by 
$\sim$15\% at 0.5 keV from that obtained with the column density used for A2052 ($2.7 \times 10^{20}$~cm$^{-2}$). Thus, we
overestimate the background model, i.e. underestimate the background-subtracted signal by $\sim$1\% at 0.5 keV for A2052.
At higher energies and for the brighter clusters the effect is smaller. Thus, the uncorrected column density variation in
the blank sky and cluster fields has a negligible effect on our results.

\subsection{\chandra ACIS}
For ACIS, we extracted background images from the standard set of cti-corrected ACIS blank sky images in the \chandra 
CALDB (Markevitch et al., 2003).  The exposure times in the background images were adjusted to produce the same 
9.0-12.0~keV count rate as in the cluster observations. Except for A85, A2029 and A2199, all observations were done
in very faint (VF) telemetry format and the VF background filtering was applied using the CIAO tool 
$acis\_process\_events$. The same VF background screening was applied to the background data sets by only including 
events with "status=0".

We estimated the effect of background modelling uncertainties (i.e. the statistical uncertainty in the 9.0-12.0~keV band
used for normalisation and the spectral variability in the band used for the cluster analysis) by varying the background
normalisation by $\pm$10\%. We propagated in quadrature the shift in the best-fit value to the parameter uncertainties.

\section{Analysis methods}
We performed an X-ray spectroscopic analysis for all the data of our cluster sample (see Sect. \ref{app_plots}). 
The temperature measurement is primarily driven by the shape of the bremsstrahlung continuum for clusters with 
kT $\gtrsim$ 2 keV. Thus, the accuracy of the temperature measurement is sensitive to the
accuracy of the calibration of the energy dependence of the effective area (see Sect. \ref{effareasimu}). The 
comparison of continuum temperatures measured with different instruments for the same cluster yields information on the 
cross-calibration accuracy of the energy dependence of the effective areas. The normalisation of the effective area 
affects the derived emission measure and flux. Thus, the comparison of cluster fluxes enables an evaluation of 
the uncertainties of the cross-calibration of the normalisation of the effective areas.

\subsection{Spectral fits}
For the spectral analysis, we modelled the emission with a single-temperature MEKAL model (Kaastra 1992). We used the 
metal abundances of Grevesse \& Sauval (1998). We used the optical measurements of cluster galaxies found from 
the NASA Extragalactic Database for the cluster redshift. To model the effect of the Galactic 
absorption, we applied the PHABS model. We fixed the column density to the value obtained from 21 cm radio observations (Kalberla et 
al., 2005) and used the absorption cross sections of Balucinska-Church \& McCammon (1992).

We binned the spectra to contain a minimum of 100 counts per bin. We fitted the background-subtracted (see 
Sect. \ref{background}) spectra in different energy bands: 0.5--7.0 keV ({\it wide band}) , 
0.5--2.0 keV ({\it soft band}) and  2.0--7.0 keV ({\it hard band}). We allowed the temperature, metal abundance and the 
model normalisation to vary in the fits, unless stated otherwise. We used the best-fit emission measures to determine the 
cluster fluxes.

\subsubsection{Fe XXV/XXVI line ratio method}
\label{feline}
In the spectral analysis described above, the shape of the bremsstrahlung continuum dominates the temperature measurement
in broad energy bands.
However, the temperature dependence of the Fe ionisation fraction can also be used as an additional, independent temperature 
measurement. In practise, the Fe XXV/XXVI line flux ratio is the most useful ionisation temperature measurement in hot 
clusters of galaxies. The energy resolution of EPIC and ACIS instruments is adequate for resolving these two lines.
The lines are centred at $\sim$6.7 and $\sim$7.0 keV in the rest frame of the clusters. 

We measured the ionisation temperature by fitting the spectra in a band of 0.8 keV width centred on the mean value of the
redshifted Fe line energies (= 6.85 keV / (1+z)) with a MEKAL model. To yield enough channels in this narrow 
band, we grouped the data with a requirement that each channel contains a minimum of 50 counts.

\subsection{Statistical analysis}
By comparing the spectroscopic results (temperatures (T) and fluxes (F) ) of our cluster sample we addressed the issue 
of systematic uncertainties in the calibration of the different instruments. For each cluster, and a given pair of 
instruments, we computed the difference of the temperature or the flux measurements and its statistical uncertainty, in 
terms of a fraction of the average value (f$_T$ = $\Delta$T/$<$T$>$ and $\sigma_{f_{T}}$ = $\sigma_{\Delta T}$/$<$T$>$;
f$_F$ = $\Delta$F/$<$F$>$ and $\sigma_{f_{F}}$ = $\sigma_{\Delta F}$/$<$F$>$). Using these values we calculated the 
weighted mean and its statistical uncertainty at 1$\sigma$ level.  
A systematic uncertainty in the calibration would tend to drive the mean of the distributions away from zero. 
We evaluate the significance of the deviation of the weighted mean from zero in terms of its statistical uncertainty.

\section{Minimising the number of temperature components}
For optimal calibration, the spectra of the chosen targets should be as simple as possible. We use a single-temperature 
plasma emission model in our analysis, to make a comparison between different instruments. Possible multi-temperature 
components in the ICM complicate the issue, because different instruments have different energy dependent effective areas
which will weight the cooler and hotter emission differently (e.g. Mazzotta et al., 2004). Thus, even a 100\% accurate 
calibration in all instruments 
may lead to different single-temperature measurements, if a very complicated temperature structure is present in the 
cluster. This is unlikely in our sample because it contains the most relaxed nearby clusters known in literature
and we only extract spectra near the peak of the temperature profile. We investigate in Sect. \ref{centralcool}
the effects of the central cooling and the temperature decline at large radii present in relaxed clusters.

\subsection{Central cooling}
\label{centralcool}
We aim to minimise the effect of the central cooling in our cluster sample by simply excluding the emission from
the cool cores in our analysis. Vikhlinin et al. (2005,2006) showed that in most clusters  
the central cooling is confined within the central 70 kpc. At the redshifts of our cluster sample, 70 kpc corresponds to 
angular diameters of 0.8--1.7$'$. We constrained the inner radius of our spectral extraction regions to be greater 
than 70 kpc.\footnote{For the additional constrain for r$_{\rm in}$, arising from the PSF, see Sect. \ref{psf}} 
A3571 and Coma are exceptions to this rule, because they do not exhibit any central cooling. We excluded the central 1$'$ region 
of Coma to avoid the contribution from the central galaxy NGC4874.

\subsection{Radial temperature gradient}
\begin{figure}
\includegraphics[width=9.5cm]{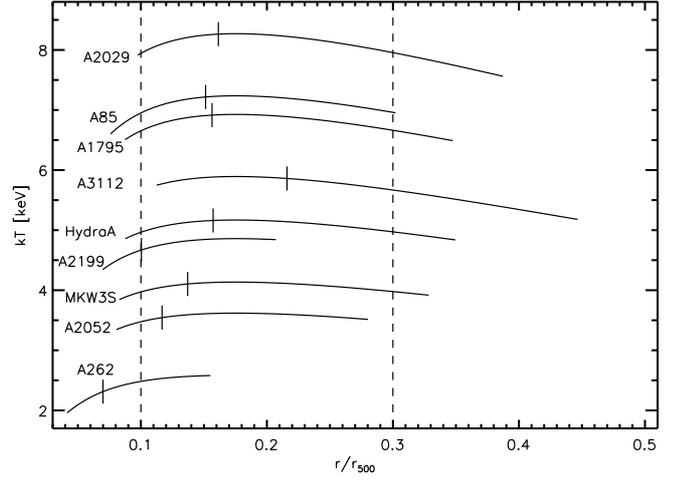}
\caption{The solid lines show the average 3D temperature profiles from Vikhlinin et al. (2006) for each cool core cluster 
in our sample. The models are plotted as a function of the 3D radius in units 
of r$_{500}$,  starting from the radii corresponding to r$_{\rm in}$ when projected to the 
plane of the sky, extending to r$_{\rm out}$ used for extracting the pn/MOS spectra. The solid vertical bars indicate r$_{\rm out}$ 
used for extracting the spectra for EPIC/ACIS comparison. The dashed vertical lines denote the approximately isothermal
region of 0.1--0.3 r$_{500}$. 
\label{tprof2.fig}}
\end{figure}

Most clusters feature a radial temperature profile that decreases at large radii, as found, e.g. from the \chandra analysis 
by Vikhlinin et al. (2005,2006). Over the 3-dimensional radial range of $0.1-0.3 r_{500}$ the cluster temperature 
profiles are rather flat, i.e. they vary by less than 15\%. We estimated the value of r$_{500}$ for our sample by using the 
T - r$_{500}$ scaling relation of Vikhlinin et al. (2006) and approximating the maximum temperature T$_{peak}$ using our 
single-temperature pn fits in the 2.0--7.0 keV band (see Table \ref{clustinfo3.tab}). For the cool core clusters, we 
further calculated the 3D temperature profile using the model for the sample average in Vikhlinin et al. (2006). The 
calculations showed that, except for the nearest cluster A262, most of the radial ranges used for the spectrum extraction 
in our sample are within the rather isothermal range of $\sim$0.1--0.3 r$_{500}$ (see Fig. \ref{tprof2.fig}).

In order to characterise the gas density distribution, we constructed surface brightness profiles using ACIS data,
except that for Coma we used the published results from ROSAT analysis (Briel et al., 1992).
We fitted the profiles with a single-$\beta$ or double-$\beta$ 
model
\begin{equation}
I(b) = I_{0,1} \times \left[1 + {\left( \frac{b}{r_\mathrm{core,1}} \right)}^2\right]^{(-3\beta_{1} + \frac{1}{2})} + I_{0,2} \times \left[1 + {\left( \frac{b}{r_\mathrm{core,2}} \right)}^2\right]^{(-3\beta_{2} + \frac{1}{2})},
\label{betamodel}
\end{equation}
where b is the projected radius. We used the best-fit parameters (see Table \ref{clustinfo3.tab}) to distribute the gas 
density with radius. We divided the cluster in concentric spherical shells and assigned each shell with a density and 
temperature given by the above models. We then intersected the spherical shells with a hollow cylinder with the inner 
and outer radii given in Table \ref{clustinfo3.tab}, which represent the projected spectral extraction region,
and computed the intersecting volumes of each shell. Finally, we calculated the emission measure of each shell at 
different temperatures along the line of sight.

We found that 4--17\% of the emission in the regions used for the EPIC/ACIS comparison originates from outside the 
nearly isothermal region (i.e. from 0.3--1.0 r$_{500}$, see Table \ref{clustinfo3.tab}). A3112 has the largest fraction (17\%) 
of its projected emission originating from the non-isothermal region and we thus used it as a conservative example on the
effect of the possible temperature bias due to the temperature gradients in our sample. We used the above 
projected temperatures and emission measures to form a composite spectrum model and used it to simulate data for pn, MOS1
and MOS2 and ACIS-S using XSPEC. We fitted the simulated isothermal emission (originating from 0.1--0.3 r$_{500}$ ) and 
the total emission  (originating from 0.1--1.0 r$_{500}$) with a single-temperature MEKAL model. The exercise showed that 
the effect of including the emission from 0.3--1.0 r$_{500}$ was to reduce the measured temperatures by $\sim$3\%.
Importantly, all instruments yielded the same temperature in all energy bands 
within $\sim$0.1\% when we included the emission from
0.3--1.0 r$_{500}$. Thus, in our sample, the projected lower temperature emission due to temperature gradient 
does not introduce a significant bias in the derived temperatures between different instruments or energy bands.

\subsection{PSF scatter}
\label{psf}
A complication when comparing \chandra and \xmm spectra is that the point spread function (PSF) of the pn 
(FWHM $\sim 6''$) and MOS (FWHM $\sim 5''$) (see  \xmm Users' 
Handbook\footnote{http://xmm.esa.int/external/xmm\_user\_support/documentation/uhb/})
is larger than that of ACIS. The tail in the King-profile shape of the EPIC PSF results in 90\% encircled 
radius of $\sim$40--50$''$, while the corresponding value for ACIS is 1--2$''$ in the 0.5-7.0 keV band (see  The \chandra 
Proposers' Observatory Guide\footnote{http://cxc.harvard.edu/proposer/POG/html/index.html}). Thus, when a bright cool 
core is present in a cluster, the tail of the PSF distribution may scatter a significant fraction of the central emission into the 
regions we examine in this work. This effect is greater in EPIC due to its wider PSF, and this may introduce 
a bias towards lower temperatures measured with the EPIC instruments.
 
To minimise the effect of scattered emission from the bright cool core, we additionally required that the inner 
extraction radius for each cluster must be greater than 1.5 arcmin (i.e. $\sim$2 times the 90\% encircled radius) for the 
cool core clusters. The inner radius of the spectral extraction region was therefore set to the maximum between the cool 
core exclusion radius (see Sect.~\ref{centralcool}) and the PSF core exclusion radius defined above except for the 
clusters without the cool core. We then estimated the contribution of the PSF-scattered flux within our adopted spectral 
extraction regions. We performed these calculations for the pn, which has the greatest effect. We used analytical expressions 
for the PSF profile of the pn from Ghizzardi (2002).
 The weak energy dependence of the PSF produces a negligible effect in the 
0.5--7.0 keV band and we thus report the contributions assuming that all photons have an energy of 2 keV. 

We convolved the \chandra surface brightness model (see Table \ref{clustinfo3.tab}) from a given input radial region with 
the pn PSF profile and calculated its relative contribution to the flux within a given output radii for pn.  The 
calculations showed that the contribution from the cool inner 70 kpc radius 
to the total emission projected within r$_{\rm in}$ and r$_{\rm out}$ ranges between 0.3\% and 6\% (see 
Table \ref{clustinfo3.tab}).

To obtain a conservative estimate for the uncorrected PSF effect, we further examined  A2052, which has the greatest 
PSF scattering contribution. We used the published \chandra temperature profile of A2052 (Blanton et al., 2003) to approximate the 
temperature within our projected radii r$_{\rm in}$ and r$_{\rm out}$ ($\sim$3.5 keV) and within the central 70 kpc 
($\sim$2.5 keV). Using XSPEC simulations, we then produced a composite spectrum containing two MEKAL components with the 
above temperatures in the proportion estimated by the PSF scattering calculations. We then fitted the composite spectrum with a 
single temperature MEKAL model, to approximate the effect of uncorrected PSF contribution. The spectral analysis shows 
that the effect of the PSF scattering is to reduce the best-fit pn temperature by $\sim$2\%, independent of the band 
choice used in the spectral analysis. Due to its smaller PSF, the effect is smaller for MOS and ACIS. Thus, the PSF scattering
may cause differences in the temperature measurements with different instruments by a maximum of 2\%, which is negligible 
compared to the statistical uncertainties. 

\begin{figure*}[t]
\includegraphics[width=16cm]{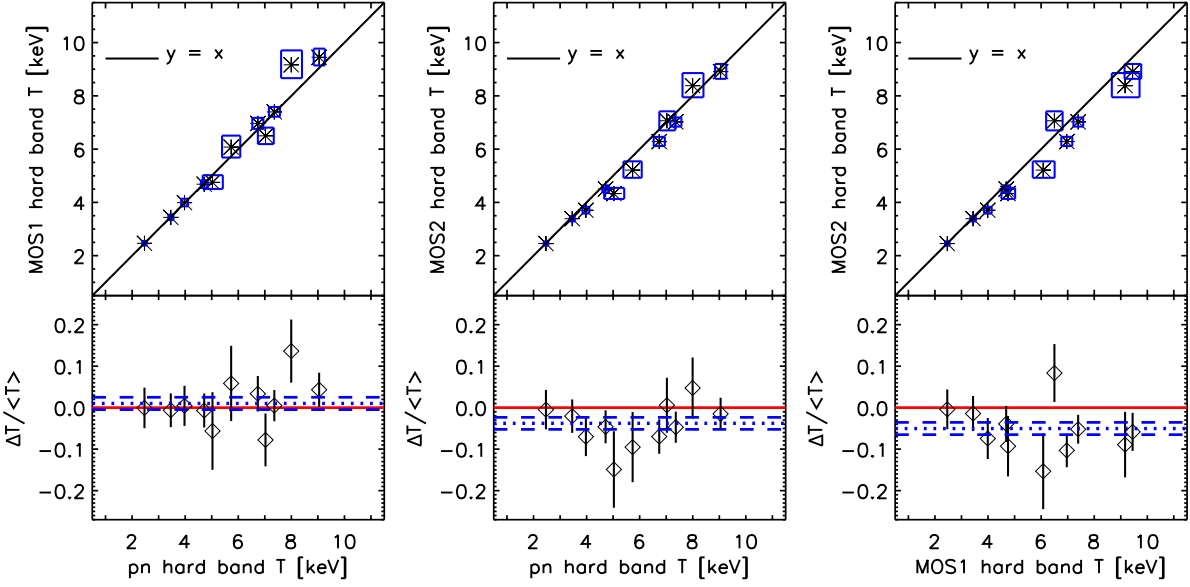}
\vspace{-2cm}
\caption{The upper panels show the best-fit temperatures (asterisks) and the 1$\sigma$ uncertainties (boxes) in the 
hard band obtained with the \xmm instruments pn (left panel) , MOS1 (middle panel) and MOS2 (right panel).
The lower panels show the relative temperature difference, $f_{T}$ = $\Delta$T/$<$T$>$ (diamonds), 
and its 1$\sigma$ uncertainty for MOS1-pn (left panel), MOS2-pn (middle panel) and MOS2-MOS1 (right panel).
The dotted and dashed lines show the weighted mean of $f_{T}$ $\pm$ the error of the mean. 
\label{pn_m1_m2_hard.fig}}
\end{figure*}

\section{Hard band temperatures}
\label{hard}
We first restricted the spectroscopic analysis to the hard band (2.0--7.0 keV). This was motivated by the fact that the 
modelling of the cluster emission is less complex because the galactic absorption does not affect the emission at these 
energies. Also, the choice of 2 keV for the break between the soft and the hard bands allows us to have some control on 
the calibration of the different instrumental components. The EPIC optical filter transmission is essentially 100\% in 
the hard band, while it is significantly 
smaller in the soft band. The molecular contamination on the ACIS filters (see below) also affects only the lowest energies. The 
quantum efficiency of the pn is essentially 100\% in the hard band, and thus the effective area of the pn primarily depends
on the effective area of the mirrors.
For MOS and ACIS, the quantum efficiency is less than 100\% in the hard  band and the different components cannot be 
distinguished.

\begin{figure*}
\includegraphics[width=13cm]{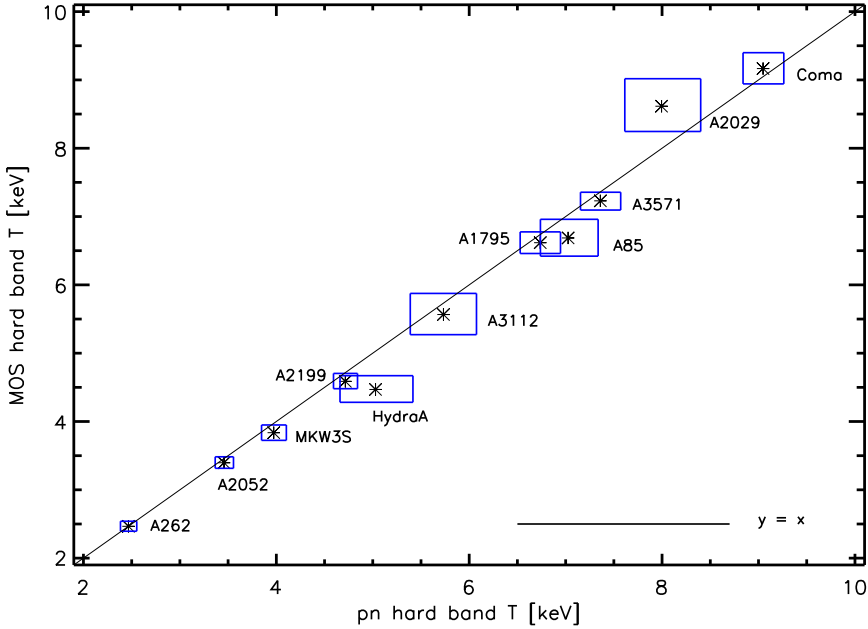}
\caption{The best-fit temperatures (asterisks) and the 1$\sigma$ uncertainties (boxes) in the hard band obtained 
with the pn v.s. those obtained with MOS (i.e. combined MOS1 and MOS2 spectra).
\label{pn_mos_hard.fig}}
\end{figure*}

\subsection{EPIC temperatures}
\label{pnmos}

\begin{table}
\caption{\label{pn_mos_hard.tab}Best-fit hard band temperatures and 1$\sigma$ uncertainties for EPIC instruments.}
\centering
\scriptsize{
\begin{tabular}{lcccccc}
\hline\hline
        &          &          & MOS1             & MOS2             & MOS              & pn  \\             
name    & r$_\mathrm{in}$ & r$_\mathrm{out}$ & T$_\mathrm{hard}$        & T$_\mathrm{hard}$       & T$_\mathrm{hard}$        & T$_\mathrm{hard}$ \\
        & [']     & [']       & [keV]            & [keV]            & [keV]            & [keV]      \\
\hline
A1795   & 1.5      & 6.0      & 7.0[6.8--7.2] & 6.3[6.1--6.5] & 6.6[6.5--6.8] & 6.7[6.5--6.9]  \\
A2029   & 1.5      & 6.0      & 9.2[8.7--9.7] & 8.4[7.9--8.9] & 8.6[8.2--9.0] & 8.0[7.6--8.4]  \\
A2052   & 1.7      & 6.0      & 3.4[3.3--3.5] & 3.4[3.3--3.5] & 3.4[3.3--3.5] & 3.5[3.4--3.6]  \\
A2199   & 2.0      & 6.0      & 4.7[4.5--4.8] & 4.5[4.4--4.6] & 4.6[4.5--4.7] & 4.7[4.6--4.8]  \\
A262    & 1.6      & 6.0      & 2.5[2.4--2.6] & 2.5[2.4--2.5] & 2.5[2.4--2.5] & 2.5[2.4--2.6]  \\
A3112   & 1.5      & 6.0      & 6.1[5.7--6.5] & 5.2[4.9--5.5] & 5.6[5.3--5.9] & 5.7[5.4--6.1]  \\
A3571   & 0.0      & 6.0      & 7.4[7.2--7.6] & 7.0[6.8--7.2] & 7.2[7.1--7.4] & 7.4[7.2--7.6]  \\
A85     & 1.5      & 6.0      & 6.5[6.2--6.8] & 7.1[6.7--7.4] & 6.7[6.4--7.0] & 7.0[6.7--7.3]  \\
Coma    & 1.0      & 6.0      & 9.4[9.1--9.8] & 8.9[8.6--9.2] & 9.2[8.9--9.4] & 9.0[8.8--9.3]  \\
HydraA  & 1.5      & 6.0      & 4.8[4.5--5.0] & 4.3[4.1--4.5] & 4.5[4.3--4.7] & 5.0[4.7--5.4]  \\
MKW3S   & 1.5      & 6.0      & 4.0[3.9--4.1] & 3.7[3.6--3.8] & 3.8[3.7--3.9] & 4.0[3.8--4.1]  \\
\hline
\end{tabular}}
\tablefoot{
Spectra were extracted from concentric annuli of inner and outer radii given as r$_{\rm in}$ and r$_{\rm out}$. 
The data were fitted with a single-temperature MEKAL model in the hard band (2--7 keV) with the absorption fixed
at the Galactic value and the abundance allowed to vary.
}
\end{table}

The single temperature MEKAL hard band fits to the pn, MOS1 and MOS2 data revealed that the pn and MOS1 values differ 
by only $\sim$1\% on average (see Fig. \ref{pn_m1_m2_hard.fig} and Table \ref{pn_mos_hard.tab}). 
There is an indication that the MOS2 temperatures are systematically lower (by $\sim$4\% and  $\sim$5\%) than those obtained with the pn 
and MOS1, respectively. However, these differences are significant only at a modest level of 2.6$\sigma$ and 3.4$\sigma$.
The combined MOS data (MOS1 + MOS2) yields $\sim$2\% lower hard band temperatures
than the pn on average and the values agree within the uncertainties at the 1 $\sigma$ confidence level (see 
Fig. \ref{pn_mos_hard.fig}). Thus, the cross-calibration of the energy dependence of the effective area 
between the pn and MOS is consistent within the statistical precision of these measurements,  which corresponds to less 
than $\sim$5\% discrepancy in temperature. We will use the pn data in the following discussion for comparisons with the 
hard band temperatures obtained with other satellites, due to the smaller statistical uncertainties with respect to the MOS 
cameras.

\subsection{\xmm v.s. \sax temperatures}
\label{xmm-bepposax}

\begin{figure}[t]
\includegraphics[width=13cm]{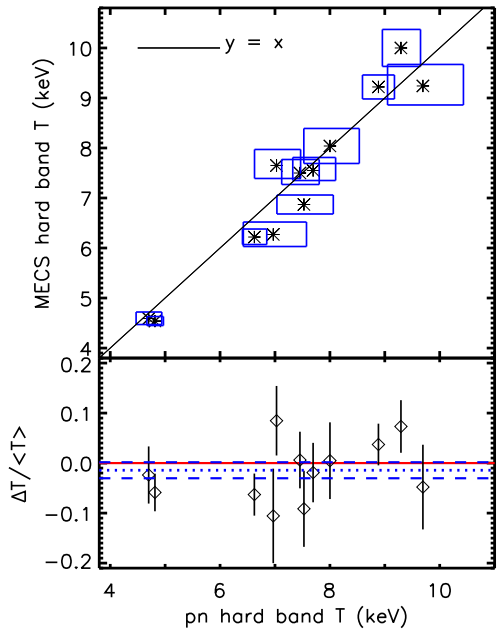}
\caption{The upper panel shows the best-fit temperatures (asterisks) and 1$\sigma$ uncertainties (boxes) in the 
hard band obtained with the \xmm pn and \sax MECS instruments.
The lower panel shows the relative MECS-pn temperature difference, $f_{T}$ = $\Delta$T/$<$T$>$ (diamonds), 
and its 1 $\sigma$ uncertainty.
The dotted and dashed lines show the weighted mean of $f_{T}$ $\pm$ the error of the mean. 
\label{pn_mecs.fig}}
\end{figure}

We found that the pn hard band temperatures only differ by $\sim$1\% (0.9$\sigma$) from the \sax MECS values on average (see 
Fig. \ref{pn_mecs.fig} and Table \ref{t_pn_mecs.tab}). The pn values are not systematically higher or lower than the MECS
values.

\subsection{\xmm v.s. \chandra temperatures}

\begin{figure*}
\includegraphics[width=15cm]{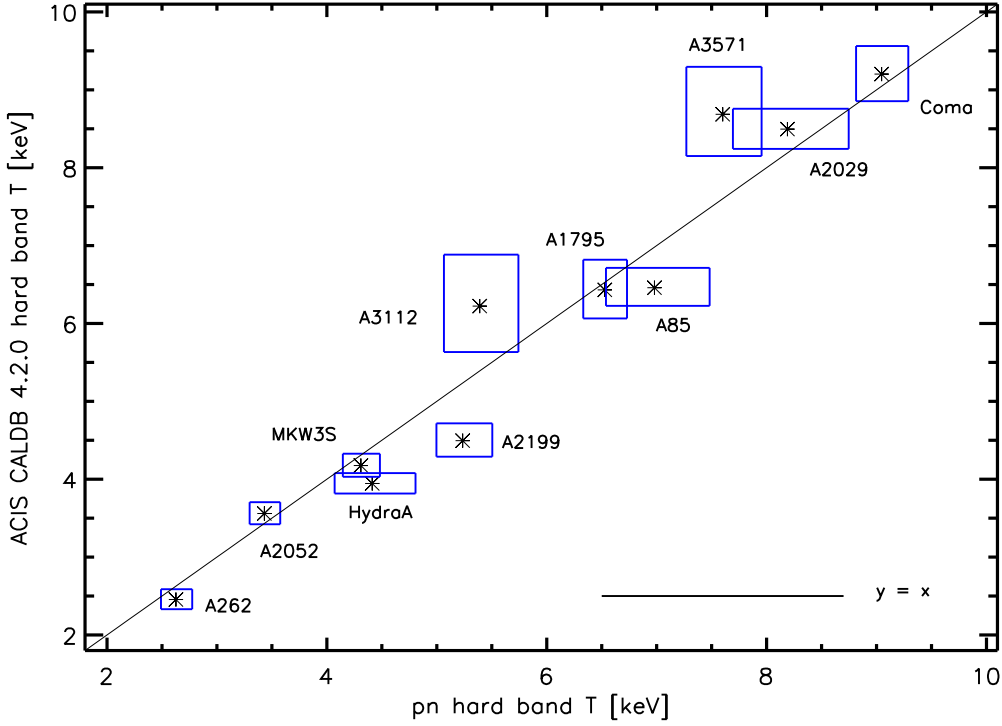}
\caption{The best-fit temperatures (asterisks) and the 1$\sigma$ uncertainties (boxes) in the hard band obtained 
with the \xmm pn detector v.s. those obtained with the \chandra ACIS detector using CALDB 4.2.0.
\label{pn_acis_hard2.fig}}
\end{figure*}

Our spectral fits to the pn and ACIS data using CALDB 4.2.0 (see Figs. \ref{pn_hardplot2.fig} and 
\ref{acis_hardplot2.fig}) showed that the ACIS hard band temperatures differ from the pn values only by $\sim$1\% 
(0.6$\sigma$) on average (see Fig. \ref{pn_acis_hard2.fig} and Table \ref{pn_acis.tab}) implying similar accuracy for 
the cross-calibration of the energy dependence of the effective area 
between ACIS and EPIC in the hard band.

\begin{table}
\caption{\label{t_pn_mecs.tab}Best-fit hard band temperatures and 1$\sigma$ uncertainties for pn and MECS.}
\centering
\begin{tabular}{lcccc}
\hline\hline
              &          &          & pn               & MECS      \\ 
name           & r$_\mathrm{in}$ & r$_\mathrm{out}$ & T$_\mathrm{hard}$        & T$_\mathrm{hard}$ \\
               & [']     & [']       & [keV]            & [keV]      \\
\hline
A1795          & 2.0      & 4.0      & 6.6[6.4--6.9]  &  6.2[6.1--6.4]   \\
A1795          & 4.0      & 6.0      & 7.0[6.4--7.6]  &  6.3[6.0--6.5]   \\
A2029          & 2.0      & 4.0      & 8.0[7.5--8.5]  &  8.0[7.7--8.4]   \\
A2199          & 2.0      & 4.0      & 4.8[4.7--5.0]  &  4.5[4.5--4.6]   \\
A2199          & 4.0      & 6.0      & 4.7[4.5--4.9]  &  4.6[4.5--4.7]   \\
A3571          & 0.0      & 2.0      & 7.5[7.1--7.8]  &  7.5[7.3--7.8]   \\
A3571          & 2.0      & 4.0      & 7.7[7.3--8.1]  &  7.6[7.4--7.8]   \\
A3571          & 4.0      & 6.0      & 7.0[6.6--7.5]  &  7.7[7.4--8.0]   \\
A85            & 2.0      & 4.0      & 7.5[7.0--8.1]  &  6.9[6.7--7.1]   \\
Coma           & 0.0      & 2.0      & 9.7[9.1--10.4] &  9.2[8.9--9.7]   \\
Coma           & 2.0      & 4.0      & 9.3[9.0--9.7]  &  10.0[9.6--10.4] \\
Coma           & 4.0      & 6.0      & 8.9[8.6--9.2]  &  9.2[9.0--9.5]  \\
\hline
\end{tabular}
\tablefoot{
The values are obtained using data extracted from concentric annuli of inner and outer radii given as r$_{\rm in}$ and 
r$_{out}$. The data were fitted with a single-temperature MEKAL model in the hard band (pn: 2--7 keV, MECS:2--10 keV).
}
\end{table}

\section{Fe XXV/XXVI line ratio}
In this Chapter, we explore the Fe XXV/XXVI line ratio method for the determination of gas temperatures.

\subsection{Statistical issues}
\label{fe_stat}
We obtain the Fe XXV/XXVI line ratio temperatures by fitting the data with the MEKAL model in the $\sim$6--7 keV 
energy band, depending on the cluster redshift (as explained in Sect. \ref{feline}). We first attempted to estimate 
whether the accuracy of the method depends on the number of photons available for the analysis. For this, we performed 
a spectral analysis of data simulated with the MEKAL model in XSPEC. We used the pn energy redistribution and auxiliary 
files of A1795 for the simulations. We used a range of 5--10 keV for the temperatures in the simulations and kept the 
metal abundance at 0.3 Solar and a redshift of zero in the simulations. We varied the number of input counts in a 
6.45--7.25~keV energy band by scaling the input model accordingly. For each temperature and number of counts, we performed 
1000 simulations, including the scatter due to counting statistics.

\begin{table}
\caption{\label{sys.tab}Systematic effects on the temperature measurement.}
\centering
\begin{tabular}{lcl}
\hline\hline
source     & $\sigma_{T}$/T    & comments                                 \\
\hline
           &                   &                       \\
\multicolumn{3}{c}{\bf MEKAL\tablefootmark{a}}\\
background & $<$ 2\%   & EPIC and ACIS, propagated     \\
PSF        & $<<$ 2\%  & EPIC and ACIS, not propagated \\
T gradient & $<$0.1\%  & EPIC and ACIS, not propagated \\
\hline
$\Sigma$   & $<2$\%    & \\
\hline
           &           &                                          \\
\multicolumn{3}{c}{\bf Fe XXV/XXVI\tablefootmark{b}}\\
limited counts     & $<$2\%  & EPIC \\
energy resolution & 2\% & pn only \\
gain       & 1--3\% & EPIC \\
MEKAL/APEC & 2\%    & EPIC \\
\hline
$\Sigma$   & 4\%    & propagated \\
\hline
\end{tabular}
\tablefoot{
\tablefoottext{a}{Temperature measurements using the MEKAL model in the hard band}
\tablefoottext{b}{Temperature measurements using the Fe XXV/XXVI line ratio}}
\end{table}

\begin{figure}
\includegraphics[width=9.3cm]{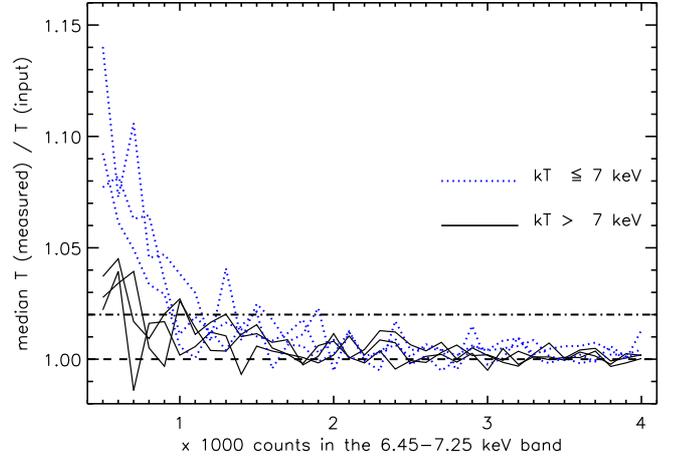}
\caption{The relative bias of the temperature derived by fitting the simulated spectra in the 6.45--7.25 
keV band as a function of the number of counts for the input values kT=5--7 keV (dotted blue lines) and 
for kT=8--10 keV (solid lines). The dash-dot line shows the allowed upper limit of 2\% for the bias in the sample.
\label{simu.fig}}
\end{figure}

We binned the simulated spectra to contain a minimum of 50 counts in each channel, as we do when fitting the actual data.
We then fitted each of the simulated spectra in the 6.45--7.25 keV band and thus obtained a distribution of best-fit 
temperatures for each input temperature and number of counts. We used these distributions to determine the median 
temperatures and compared them with the input values. The comparison showed that while the derived temperatures agreed 
with the input values within 1\% at the highest numbers of counts ($>$ 3000), at the low count numbers there is a 
tendency towards higher temperatures, especially for the lowest input temperatures (see Fig. \ref{simu.fig}). This effect 
was identical regardless of whether we used $\chi^2$ or C-statistics. At the 
lowest temperatures, the Fe XXVI line emission 
is quite weak and it is possible that the Fe XXVI line-like features are created at random due to low statistics. These 
artificial features are then interpreted as real Fe XXVI emission and the temperatures are biased high, as measured 
above.

We adopt as a selection criterium for our sample that the uncertainty due to the limited number of counts biases the 
temperature measurements by less than 2\%. We consider this 2\% uncertainty below when discussing the various 
sources of uncertainty in the temperature measurements. For clusters with temperature above $\sim$7 keV, this requirement 
imposes a lower limit of $\sim$1000 counts in the 6.45--7.25 keV band. For cooler clusters, the requirement for the 
minimum number of counts is higher. 

These limits are quite severe when compared to the observed numbers of counts in our sample. This is especially true for 
the EPIC--ACIS comparison where the inner extraction radius is limited due to our minimisation of PSF scattering, and the
outer extraction radius is limited by the ACIS FOV. Consequently, the number of counts in the $\sim$6--7 keV 
band in most clusters is well below the above limits. Thus, with the current exposures we cannot perform a statistically 
meaningful comparison of Fe XXV/XXVI temperatures between EPIC and ACIS. On the other hand, due to the larger 
regions used in the pn--MOS comparison, the number of counts in the $\sim$6--7 keV band exceeds 1000 in most clusters. Out of 
these clusters, we chose those with kT$>$ 5 keV for our analysis, i.e. A1795, A2029, A3112, A3571, A85 and 
Coma.

\begin{figure*}
\includegraphics[width=15cm]{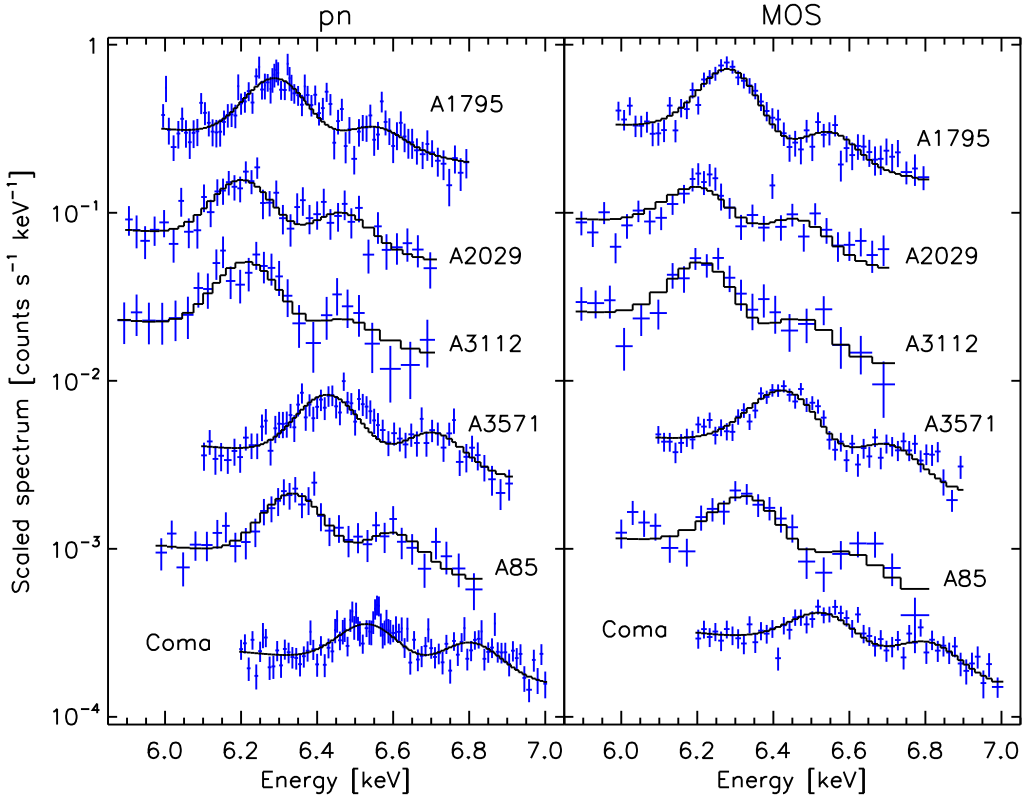}
\caption{The best-fit MEKAL models (solid line) in the $\sim$6--7 keV band data which contain the Fe XXV and Fe XXVI
emission lines and statistical uncertainties (crosses). The pn (left panel) 
and MOS (right panel) data and the models are arbitrarily scaled for the clarity of the display. 
\label{6070plot.fig}}
\end{figure*}

\subsection{Systematic uncertainties}
The accuracy of the Fe XXV/XXVI line ratio measurement additionally depends on the accuracy of the 
absolute energy scale and the energy resolution used to generate the 
redistribution matrix in the $\sim$6--7 keV band. 
Also, uncertainties in the redshift measurement and the atom physics used to generate the MEKAL model
have some effect to the interpreted temperature for a 
given line ratio data. We found that these effects (see Appendix \ref{sys_app} for the details), combined with the bias 
due to limited number of counts (Sect. \ref{fe_stat}) may affect the EPIC temperatures at the $\sim$4\% level (see 
Table \ref{sys.tab}). We will consider this uncertainty and propagate it along with the statistical uncertainties of EPIC 
Fe XXV/XXVI temperatures when appropriate in the following.

\subsection{Results}
\label{emconstr}
We then fitted the EPIC data of the subsample described in Sect. \ref{fe_stat} (see Fig. \ref{6070plot.fig})
in the $\sim$6--7 keV band with a single-temperature MEKAL model.
To obtain a similar level of counts as in the pn instrument, we co-added the MOS1 and MOS2 data for comparison.

\begin{figure}
\hbox{
\includegraphics[width=6.0cm,angle=0]{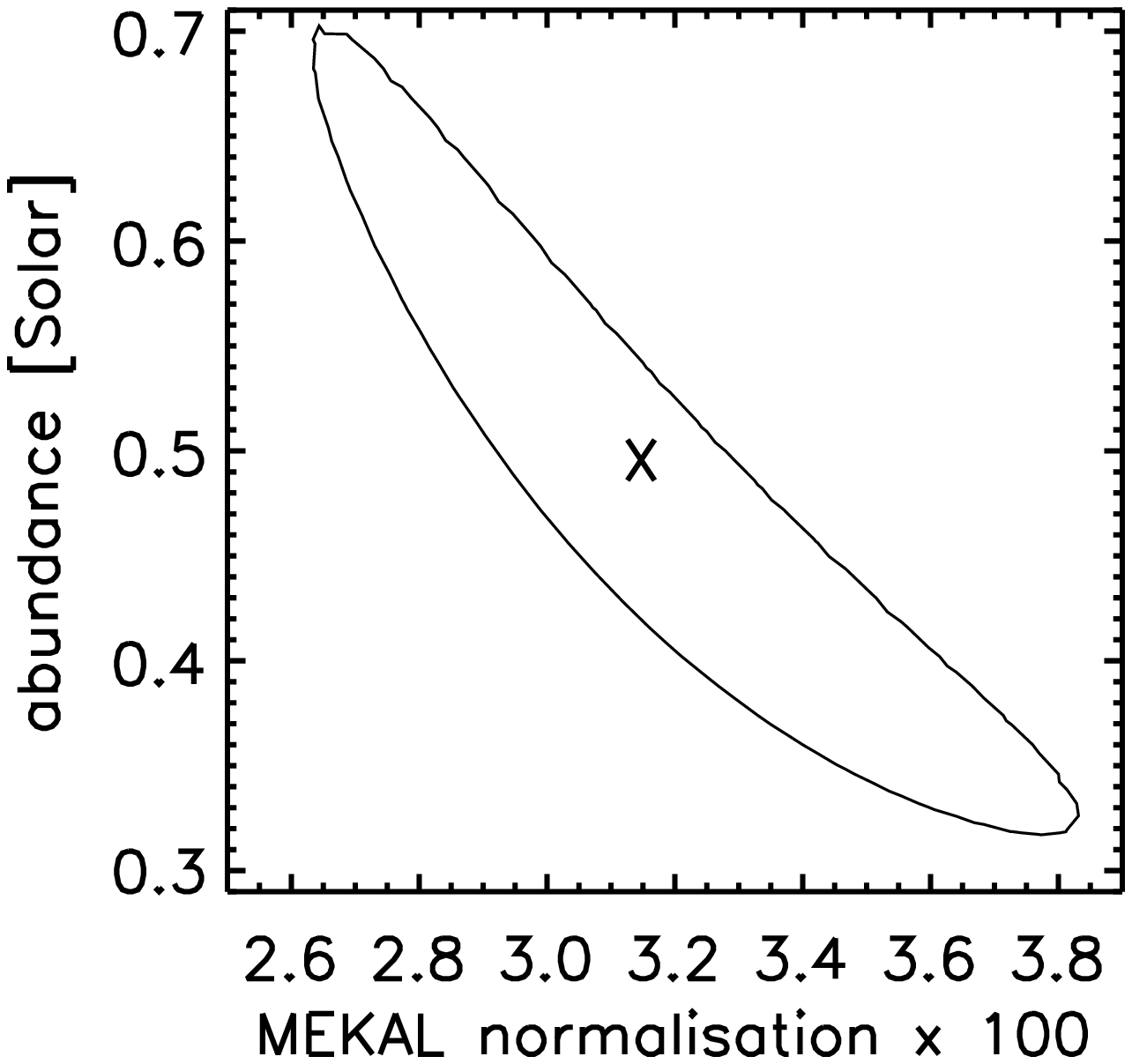}
\hspace{-1.5cm}
\includegraphics[width=6.0cm,angle=0]{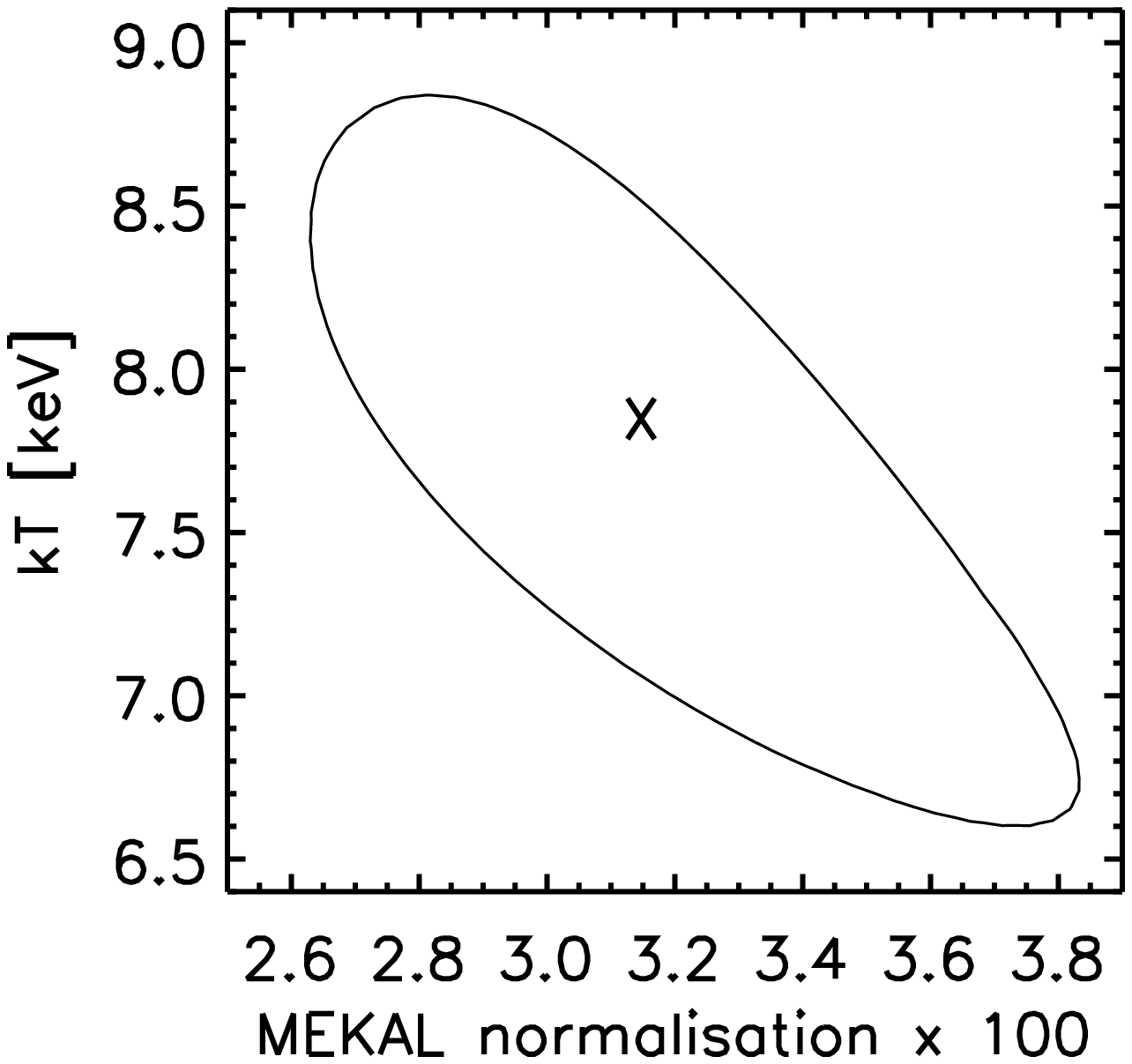}
}
\caption{The 2-parameter confidence contours at 1$\sigma$ for A2029 obtained with the MEKAL fits to 
the pn data in the $\sim$6--7 keV band.
\label{a2029_cont.fig}}
\end{figure}

The relatively small number of photons in the narrow energy band (6.45--7.25 keV in the rest frame of a cluster) 
inevitably limits the statistical accuracy of the derived temperatures. The relative 1$\sigma$ statistical uncertainty on the 
temperatures is at most 20\% (see Table \ref{pn_mos_6070.tab}). In the 6.45--7.25 keV energy 
band, the metal abundance and the emission measure are highly degenerate: a lower normalisation for the continuum is 
approximately compensated with a higher metal abundance, since the continuum cannot be well separated from the line 
profile in this narrow energy band (see Fig. \ref{a2029_cont.fig}).

\begin{table*}
\caption{\label{pn_mos_6070.tab}Best-fit temperatures and statistical uncertainties for the EPIC instruments in the 
$\sim$6--7 keV band.}
\centering
\begin{tabular}{lcc|cccc|cccc}
\hline\hline
        &          &          & \multicolumn{4}{c}{pn}  &  \multicolumn{4}{c}{MOS}       \\ 
        &          &          & \multicolumn{2}{c}{EM free\tablefootmark{a}} & \multicolumn{2}{c}{EM constr.\tablefootmark{b}}  & 
\multicolumn{2}{c}{EM free\tablefootmark{a}} & \multicolumn{2}{c}{EM constr.\tablefootmark{b}} \\
 name    & r$_{\rm in}$ & r$_{\rm out}$ & T & $\sigma_{T}$/T  &  T & $\sigma_{T}$/T & T & $\sigma_{T}$/T  &  T & $\sigma_{T}$/T \\
        & [']     & [']   &  [keV]          &        & [keV]         &      & [keV] &     & [keV]  &        \\
\hline
A1795   & 1.5      & 6.0  & 6.1[5.7--6.5] & 7\%  & 6.6[6.4--6.8] &  3\% & 5.7[5.3--6.1] &  7\% & 6.1[5.9--6.3] & 3\% \\ 
A2029   & 1.5      & 6.0  & 7.7[7.1--8.4] & 9\%  & 7.7[7.3--8.2] &  6\% & 9.0[8.3--9.7] &  8\% & 8.4[7.8--8.9] & 7\% \\ 
A3112   & 1.5      & 6.0  & 5.6[4.3--6.7] & 21\% & 5.6[5.2--6.2] &  9\% & 7.2[6.4--8.1] & 11\% & 6.4[5.9--6.9] & 8\% \\ 
A3571   & 0.0      & 6.0  & 7.5[7.0--7.9] &  6\% & 7.5[7.2--7.9] &  4\% & 6.8[6.5--7.1] &  5\% & 6.8[6.6--7.0] & 3\% \\ 
A85     & 1.5      & 6.0  & 7.4[6.6--8.1] & 10\% & 7.3[6.8--7.7] &  7\% & 6.4[5.4--7.2] & 15\% & 6.4[5.8--6.8] & 8\% \\ 
Coma    & 1.0      & 6.0  & 9.3[8.9--9.8] &  5\% & 9.0[8.7--9.3] &  4\% & 9.2[8.6--9.7] &  6\% & 8.9[8.5--9.3] & 4\% \\ 
\hline
\end{tabular}
\tablefoot{
The uncertainties are given at 1$\sigma$.
The temperatures are derived using data 
extracted from concentric annuli of inner and outer radii given as r$_{\rm in}$ and r$_{\rm out}$.
The data were fitted with a single-temperature MEKAL model in the $\sim$6--7 keV band 
\tablefoottext{a}{allowing the emission measure to be free} or  
\tablefoottext{b}{constraining the emission measure within that derived in the hard band}.
}
\end{table*}

On the other hand, we have already constrained the emission measure when fitting the hard band (2--7 keV) 
(Sect. \ref{hard}), typically within 1\% at 1$\sigma$ level. We thus experimented by constraining the emission measure 
to that found in 
the hard band fit, when fitting the $\sim$6-7 keV band. The resulting statistical uncertainties for the temperature are 
3--10\%, instead of 5--20\% in the case of a free emission measure (see 
Table \ref{pn_mos_6070.tab}). Thus, with this method, we reach the level of the systematic uncertainties in some cases
(see Table \ref{sys.tab}). In 
both cases, the temperatures are consistent within the statistical 1$\sigma$ uncertainties. Importantly, the temperatures
obtained with the free or constrained emission measure in the $\sim$6-7 keV band do not indicate a systematic difference
(see Table \ref{pn_mos_6070.tab}). This proves that the use of the hard band emission measure constrain does not introduce 
any bias to the EPIC temperature measurements.

\begin{figure}
\includegraphics[width=10cm]{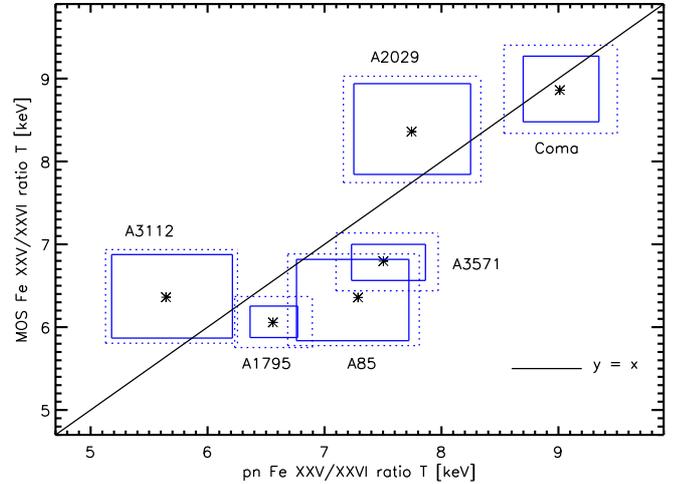}
\caption{The best-fit temperatures (asterisks) obtained with the MEKAL fits in the $\sim$6--7 keV band data
of the pn v.s. those obtained with MOS. The solid boxes indicate the statistical uncertainties at 1$\sigma$ level while the
dotted boxes include the systematic uncertainties.
\label{pn_mos_6070.fig}}
\end{figure}

The pn and MOS temperatures obtained with the hard band emission measure constrain mostly agree within the statistical 
1$\sigma$ uncertainties (see Table \ref{pn_mos_6070.tab} and Fig. \ref{pn_mos_6070.fig}). The values differ somewhat for
A1795 and A3571, which have the smallest statistical uncertainties. Considering the systematic uncertainties, the values 
agree. There is no systematic difference between the derived temperatures for the full sample obtained with the two instruments. 
Thus, in the following we will refer to the Fe XXV/XXVI temperatures obtained with the emission measure constrain.

\subsection{Fe XXV/XXVI line ratio v.s. continuum}
The $\sim$6--7 keV band temperatures above are given by the temperature dependence of the ionisation fraction, i.e. by 
the Fe XXV/XXVI line ratio. On the other hand, the hard band temperatures are primarily determined by the temperature 
dependence of the bremsstrahlung continuum. There are several reasons why the two measurements may not agree.
Due to cluster mergers, the intracluster material may be in a non-equilibrium ionisation state which would change 
the Fe XXV/XXVI line ratio from that given by ionisation equilibrium assumed in our work (e.g. Prokhorov, 2010;
Akahori \& Yoshikawa, 2010). The mergers may also accelerate a fraction of electrons into suprathermal velocities, which 
will produce deviations from a Maxwellian electron velocity distribution assumed when modelling the continuum emission
with a bremsstrahlung model (Prokhorov et al., 2009). Also, the non-Maxwellian electron distribution will affect the 
Fe XXV/XXVI line ratio (e.g. Kaastra et al., 2009). Furthermore, a relativistic electron population created by a strong merger
shock may produce an additional continuum emission component via inverse compton scatter of cosmic microwave background 
photons (e.g. Sarazin, 1988).
These effects are probably small in our sample because we used the most dynamically relaxed clusters. Also, we excluded the 
cluster outskirts, which may still be accreting material and where ionisation equilibrium may not yet have been 
established.

With the current \xmm EPIC data we are able to test the above possibilities by a direct comparison of the temperatures 
derived from both the continuum and the Fe XXV/XXVI line ratio. To keep the two temperature measurements as 
independent as possible, we derived the continuum temperature using a MEKAL fit in the 2.0--6.0 keV band. In this band, the
line emission is negligible for hot clusters and thus the metal abundance may be inaccurately determined. We examined 
this issue by fixing the metal abundance to values in the range 0.0--0.5 solar. Examining the best-fit models showed that the temperatures varied by less than 1\% for a given cluster when varying
the abundance. Thus, uncertainties in the metal abundance do not significantly bias the temperature measurements.

\begin{figure}
\includegraphics[width=10cm]{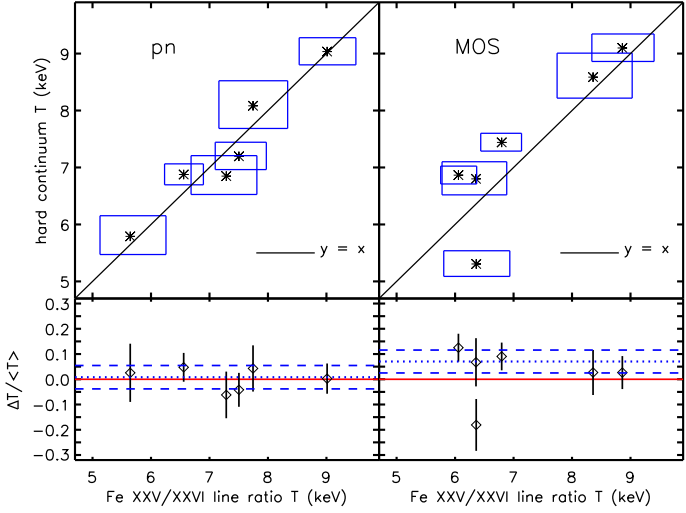}
\caption{The upper panels show the best-fit temperatures (asterisks) and 1$\sigma$ uncertainties (boxes)
for the Fe XXV/XXVI line ratio temperatures v.s. the 2--6 keV band continuum for the pn  (left panel) and MOS (right 
panel). The lower panels show the relative temperature difference $f_{T}$ (i.e. the 2--6 keV band continuum temperature 
minus the Fe XXV/XXVI line ratio temperature, divided by the mean of the two) and its 1$\sigma$ uncertainty for the pn 
(left panel) and MOS (right panel). 
The dotted and dashed lines show the weighted mean of $f_{T}$ $\pm$ the error of the mean. 
\label{pn_mos_2060_6070.fig}}
\end{figure}

The comparison of the temperatures derived by fitting the 2.0--6.0 keV and $\sim$6--7 keV band reveal a good agreement 
between the two methods (see Fig. \ref{pn_mos_2060_6070.fig}): the PN temperatures differ by $\sim$1\% (0.2$\sigma$) 
while MOS differences are  $\sim$7\% (1.6$\sigma$) 
on average. There is no systematic difference between the measurements.
This implies that the calibration of the energy dependence of the effective area of the EPIC instruments in the hard band is 
accurate and that the deviations from the ionisation equilibrium state and Maxwellian electron velocity distribution in 
these clusters are negligible. Given our evidence for the Fe XXV/XXVI line ratio measurement being insensitive to 
calibration uncertainties in the effective area (see Appendix \ref{sys_tot}), the above finding could be useful for the calibration of the 
hard band of future X-ray missions: the hard band calibration should be fine-tuned to yield the same temperature as does 
the Fe XXV/XXVI line ratio.

\section{Soft band temperatures}
\label{soft}
We then performed a spectroscopic analysis of the data in the soft band (0.5--2.0 keV). We used a single-temperature 
MEKAL model, as in the analysis of the hard band above. We allowed the temperature, metal abundance and 
emission measure to vary independently from the hard band values presented above. This was required in order to obtain 
statistically acceptable fits in the soft band (see Figs. \ref{pn_softplot2.fig} and 
\ref{acis_softplot2.fig} for pn and ACIS). Note that the formally accurate models for the soft and hard bands, which are 
adequate for calibration work, do not yield a physically consistent modelling of the full energy band. In order to keep 
the modelling and the temperature comparison simple, we excluded the coolest cluster A262 from this comparison, because 
its strong line emission at $\sim$1 keV would require multi-component modelling for an accurate description of the data.

\subsection{pn v.s. MOS}
\begin{figure}
\includegraphics[width=9cm]{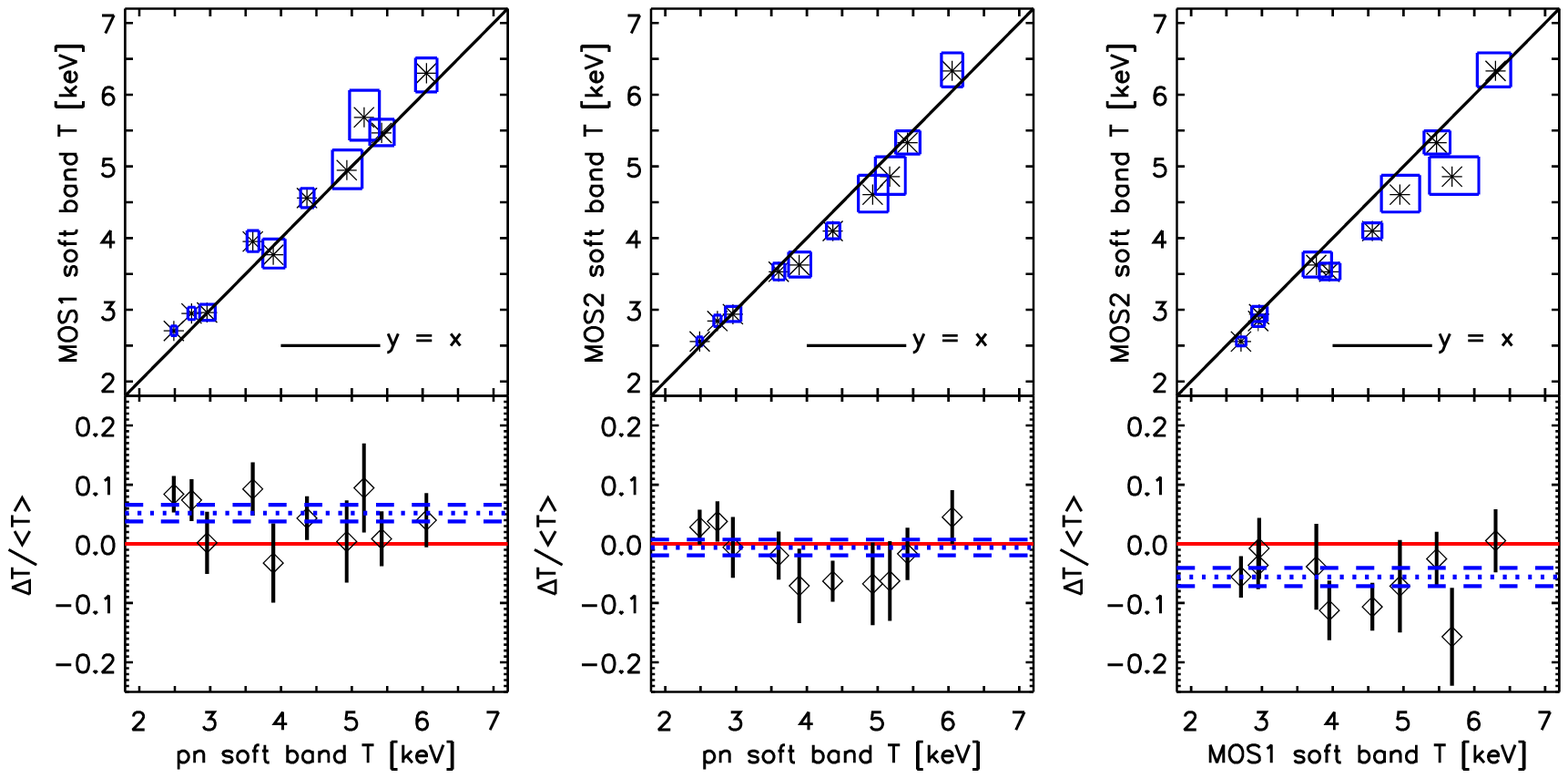}
\vspace{-2cm}
\caption{The upper panels show the best-fit temperatures (asterisks) and statistical 1$\sigma$ uncertainties (boxes) in the
soft band with the \xmm instruments pn/MOS1 (left panel), pn/MOS2 (middle panel) 
and MOS1/MOS2 (right panel). The lower panels show the relative temperature difference $f_{T}$ ( = $\Delta$T/$<$T$>$, 
diamonds) and its 1$\sigma$ uncertainty for MOS1-pn (left panel), MOS2-pn (middle panel) and MOS2-MOS1 (right panel).
The dotted and dashed lines show the weighted mean of $f_{T}$ $\pm$ the error of the mean. 
\label{pn_m1_m2_soft.fig}}
\end{figure}

We found that while the pn and MOS2 agree very well (average difference is $\sim$1\% , i.e. 0.5$\sigma$), the MOS1 
values are $\sim$5\% ($\sim$3.7 $\sigma$) higher than those of the pn and MOS1 (see Fig. \ref{pn_m1_m2_soft.fig}). 
These differences indicate some remaining cross-calibration uncertainties
in the EPIC instruments in the soft band which we will not examine in more detail here.

\subsection{pn v.s. ACIS}
\label{pn_acis_soft}
\begin{figure}
\includegraphics[width=10cm]{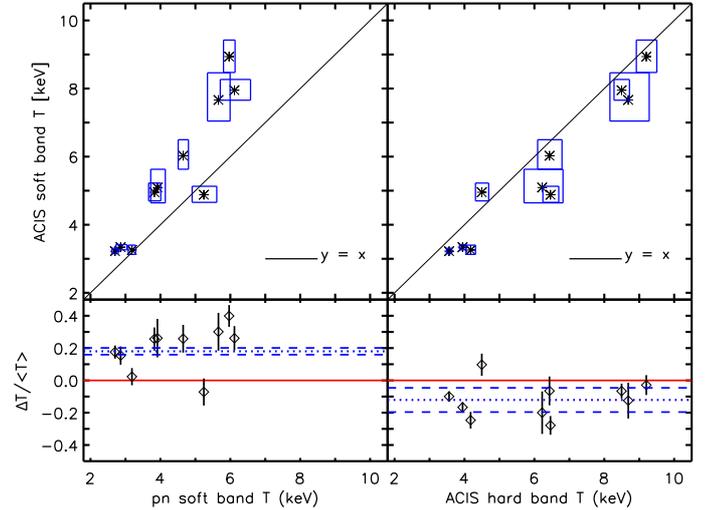}
\caption{The upper panels show the best-fit temperatures (asterisks) and 1$\sigma$ uncertainties (boxes) 
in the soft band for the pn and ACIS (left panel) and in the hard and soft bands for ACIS (right panel).
The lower panels show the relative temperature difference, $f_{T}$ = $\Delta$T/$<$T$>$ (diamonds) 
and its statistical 1$\sigma$ uncertainty for ACIS soft band - pn soft band (left panel) and ACIS soft band - ACIS hard band 
(right panel).
The dotted and dashed lines show the weighted mean of $f_{T}$ $\pm$ the error of the mean. 
\label{pn_acis_soft_hard.fig}}
\end{figure}

We found that the ACIS soft band temperatures are $\sim$18\% higher than those obtained with the pn 
(see Fig. \ref{pn_acis_soft_hard.fig}). This difference is statistically very significant, 8.6$\sigma$,
and thus there are significant remaining ACIS/pn cross-calibration uncertainties in the soft band.

To examine this difference in more detail, we convolved the best-fit soft band pn models through the ACIS responses and 
compared this prediction with the ACIS data. In the case of A85 and Coma the data are spread across several CCD chips 
and for these clusters we co-added the data and formed a single average file for the responses and background. We scaled 
the models to match the ACIS data at 2 keV, since we are primarily interested in the energy dependence of the effective area 
here, whose cross-calibration is very accurate in the 2--7 keV band (see Sect. \ref{hard}).

\begin{figure*}
\includegraphics[width=18cm]{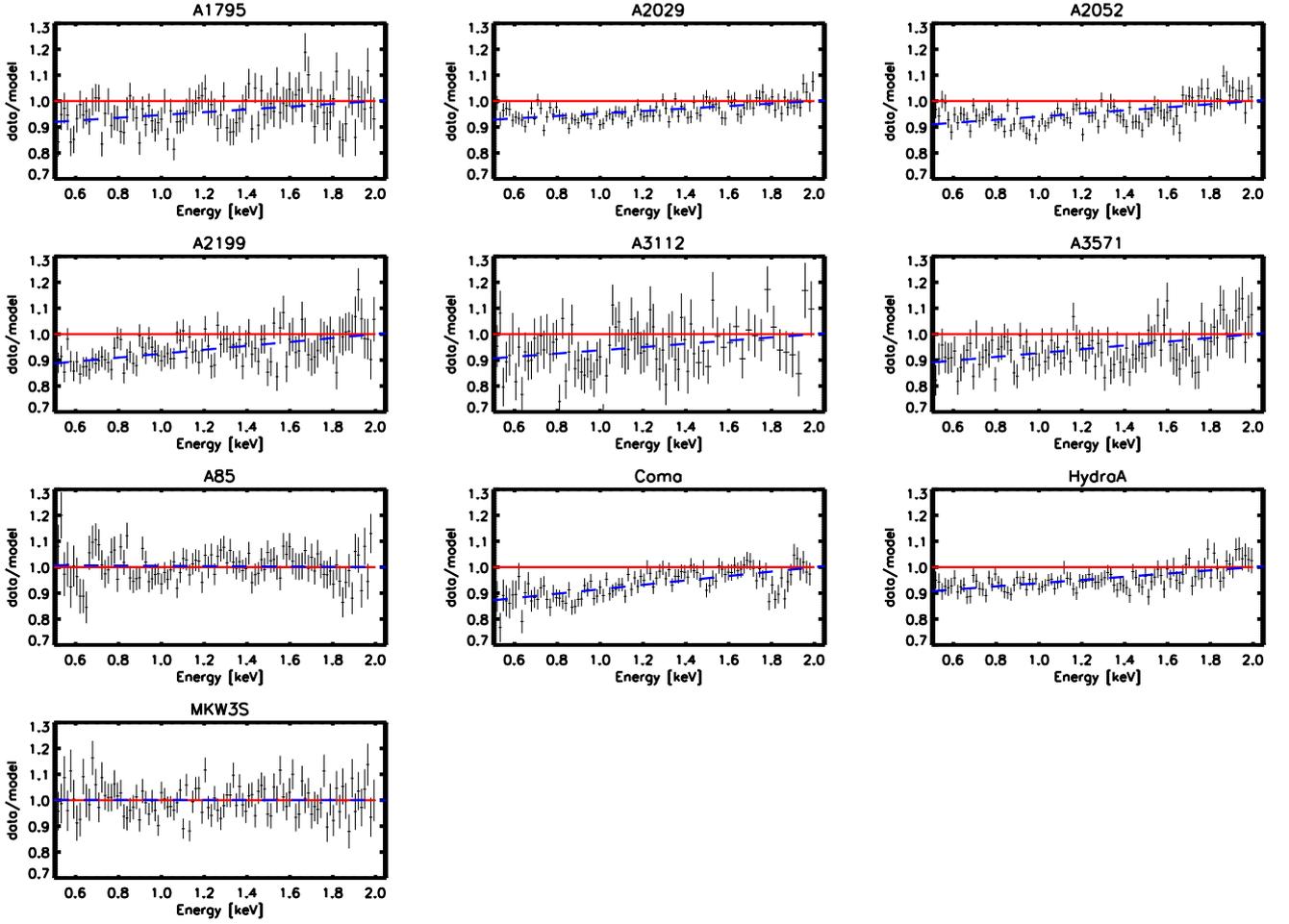}
\caption{The ratio of the ACIS data to the pn model prediction using the best-fit pn model and the ACIS responses. 
The model has been normalised so that the ratio equals unity at 2.0 keV. The dashed line shows the best-fit linear relation. 
\label{acis_pn_soft.fig}}
\end{figure*}

We found that the ratio of the ACIS data to the pn model prediction exhibits a systematic trend in most clusters.
Fitting the data-to-model ratio with a linear function we found that while the ratio is 1.0 at 2 keV by definition,  
it decreases to $\sim$0.9 at 0.5 keV in most clusters (see Fig. \ref{acis_pn_soft.fig}). This implies that
either the effective area of the pn is underestimated, or that of ACIS is overestimated at 0.5 keV by $\sim$10\%.

\begin{table}[h]
\caption{\label{pn_mos_soft.tab} The soft band temperatures and 1$\sigma$ uncertainties using EPIC instruments.} 
\centering
\begin{tabular}{lccccc}
\hline\hline
       &          &          & pn              & MOS1              & MOS2              \\             
name    & r$_\mathrm{in}$ & r$_\mathrm{out}$ & T$_\mathrm{soft}$     & T$_\mathrm{soft}$         & T$_\mathrm{soft}$       \\
        & [']     & [']       & [keV]          & [keV]             & [keV]            \\
\hline
A1795   & 1.5      & 6.0      & 4.4[4.3--4.5]  & 4.6[4.4--4.7] & 4.1[4.0--4.2] \\
A2029   & 1.5      & 6.0      & 5.2[5.0--5.4]  & 5.7[5.4--6.1] & 4.9[4.6--5.1] \\
A2052   & 1.7      & 6.0      & 2.5[2.4--2.5]  & 2.7[2.6--2.8] & 2.6[2.5--2.6] \\
A2199   & 2.0      & 6.0      & 3.6[3.5--3.7]  & 4.0[3.8--4.1] & 3.5[3.4--3.7] \\
A3112   & 1.5      & 6.0      & 3.9[3.7--4.1]  & 3.8[3.6--4.0] & 3.6[3.5--3.8] \\
A3571   & 0.0      & 6.0      & 5.4[5.3--5.6]  & 5.5[5.3--5.7] & 5.3[5.2--5.5] \\
A85     & 1.5      & 6.0      & 4.9[4.7--5.1]  & 4.9[4.7--5.2] & 4.6[4.4--4.9] \\
Coma    & 1.0      & 6.0      & 6.1[5.9--6.2]  & 6.3[6.0--6.5] & 6.3[6.1--6.6] \\
HydraA  & 1.5      & 6.0      & 3.0[2.9--3.1]  & 3.0[2.9--3.1] & 2.9[2.8--3.0] \\
MKW3S   & 1.5      & 6.0      & 2.7[2.7--2.8]  & 2.9[2.9--3.0] & 2.8[2.8--2.9] \\
\hline
\end{tabular}
\tablefoot{
The data are extracted from concentric annuli of inner and outer radii given as r$_{\rm in}$ and r$_{\rm out}$. 
The data were fitted with a single-temperature MEKAL model in the soft band (0.5--2.0 keV).
}
\end{table}

\section{Wide band temperatures}

\begin{figure}
\includegraphics[width=11cm]{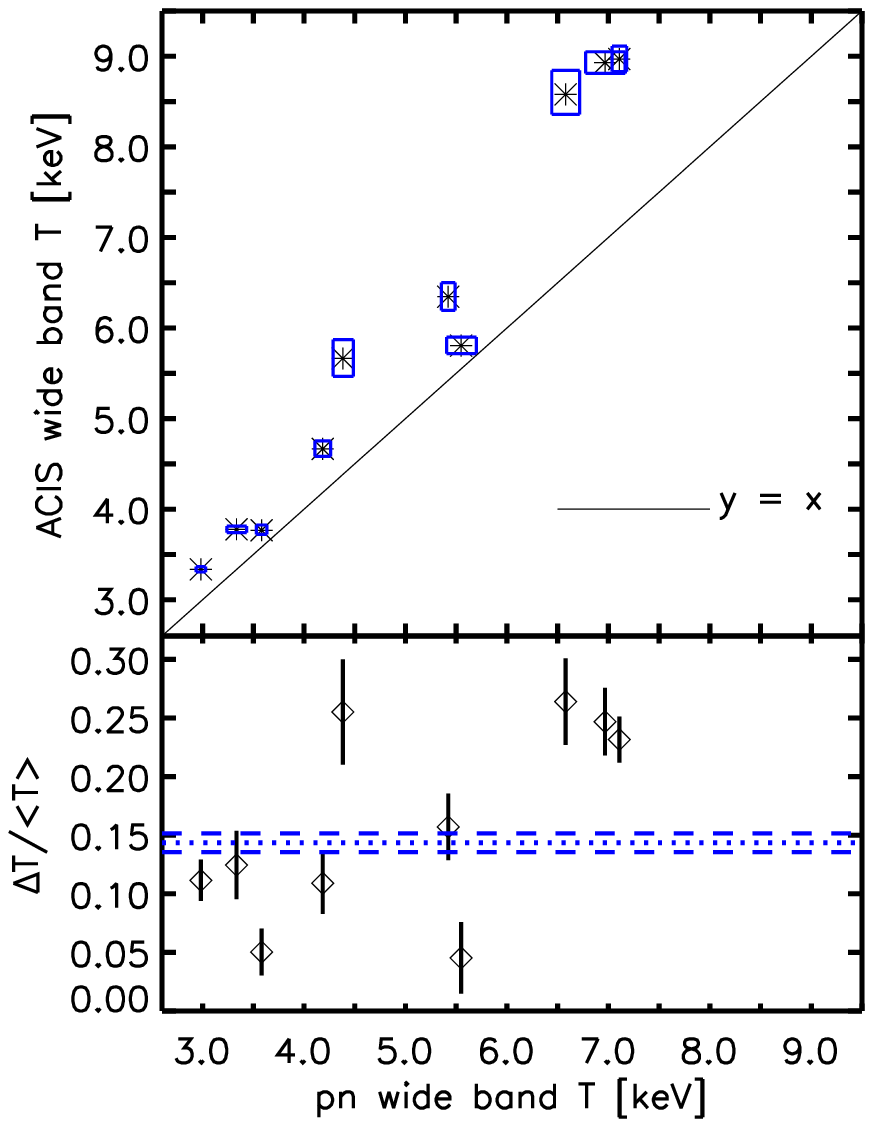}
\caption{The upper panels show the best-fit temperatures (asterisks) and 1$\sigma$ uncertainties (boxes)  
in the wide band using pn or ACIS data. The lower panel shows the relative ACIS-pn temperature difference 
$f_{T}$ ( = $\Delta$T/$<$T$>$, diamonds) and its statistical 1$\sigma$ uncertainty. 
The dotted and dashed lines show the weighted mean of $f_{T}$ $\pm$ the error of the mean. 
\label{pn_acis_wide.fig}}
\end{figure}

Since both the intrinsic cluster flux and the effective area of EPIC and ACIS are higher in the soft band, the statistical 
weight of the soft band data is much greater than that of the hard band. 
As shown above, the greatest differences between the accuracy of the EPIC and ACIS effective areas are below 2 keV which 
thus affect the temperatures derived in the wide band (0.5--7.0 keV).
For the interest of the general user, we also evaluated the effect of the cross-calibration uncertainties 
when fitting the 0.5--7.0 keV band with a single-temperature model as is often done in the 
literature. Since the hard and soft band temperatures are inconsistent in all clusters and for all instruments, the application
of a single-temperature model in the wide band will result in some residuals. These residuals are usually hidden by the 
larger statistical uncertainties in detailed temperature profile measurements presented in the literature. 

As expected, due to the above findings (Sect. \ref{pn_acis_soft}), ACIS yields systematically higher wide band 
temperatures than EPIC. On average, the ACIS/MOS1, ACIS/MOS2 and ACIS/pn differences are $\sim$8\%, $\sim$15\% and $\sim$14\% 
which corresponds to a very high detection significance of 10.0$\sigma$ 18.9$\sigma$ and 18.0$\sigma$. 
The maximum difference is $\sim$25\% (see Fig. \ref{pn_acis_wide.fig}). These values may be taken as a systematic uncertainty 
in the wide band temperature measurement when using EPIC or ACIS data. These uncertainties also apply to the analysis of EPIC 
and ACIS wide band continuum spectra of other types of objects. Evaluation of this effect remains to be carried out by analysis of e.g. 
blazars observed simultaneously with \xmm and \chandra.

\begin{figure*}
\includegraphics[width=15cm]{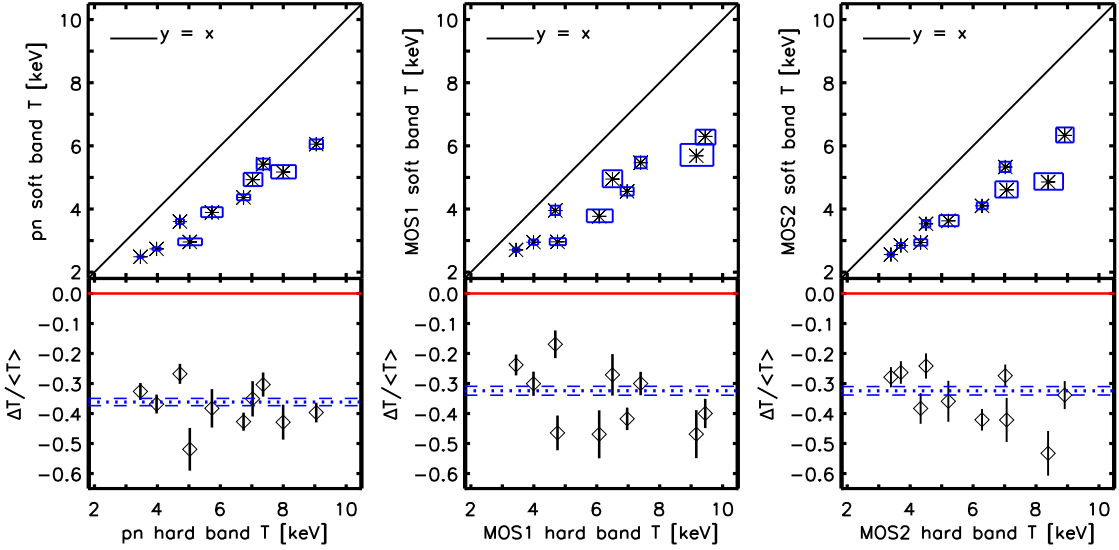}
\vspace{-3cm}
\caption{The upper panels show the best-fit temperatures (asterisks) and statistical 1$\sigma$ uncertainties (boxes) 
in the soft and hard bands for \xmm instruments pn (left panel), MOS1 (middle panel) 
and MOS2 (right panel). The lower panels show the relative soft-hard temperature difference 
$f_{T}$ ( = $\Delta$T/$<$T$>$, diamonds) and its statistical 1$\sigma$ uncertainty for pn (left panel), MOS1 (middle 
panel) and MOS2 (right panel). 
The dotted and dashed lines show the weighted mean of $f_{T}$ $\pm$ the error of the mean. 
\label{pn_m1_m2_soft_hard.fig}}
\end{figure*}

\section{Hard band v.s. soft band temperatures}

With our data set we can also compare the temperature measurements in the hard and soft bands. In 
principle, this could be an useful tool for discovering emission components in addition to the bulk of the
isothermal emission.

We found that all of the \xmm soft band temperatures are significantly smaller than the hard band values 
(see Fig. \ref{pn_m1_m2_soft_hard.fig} and Tables \ref{pn_mos_hard.tab} and \ref{pn_mos_soft.tab}): 
The pn, MOS1 and MOS2 soft band temperatures are $\sim$30--35\% ($\sim$22--30$\sigma$) lower than those obtained in the hard band. 
Also, most of the ACIS soft band temperatures are lower than the hard band values, but by a smaller amount, 
12\% (1.6$\sigma$, see Fig. \ref{pn_acis_soft_hard.fig} and Table \ref{pn_acis.tab}).  
It would be useful to examine whether this effect is present in the data of objects of different type.
While the scatter is quite large and no clear patterns can be seen, there is some indication that the relative difference
increases with temperature.

These systematic differences between the hard and soft band temperatures in all instruments could be explained by 
deviations from the isothermal emission. Since we have minimised the gas cooling effect,  
among the candidates for the excess emission are multi-temperature gas due to, e.g., clumping or 
cluster mergers and the Inverse Compton scatter of Cosmic Microwave Background photons from relativistic cluster electrons 
(see Durret et al. 2008 and Rephaeli et al. 2008 for recent reviews). Typically this excess emission flux is  
$\sim$10\% of the isothermal component. We found above that there are remaining uncertainties in the 
calibration of the effective area in the studied instruments at this level in the soft band. 
Due to these possible astrophysical effects in the clusters of galaxies, they are not a good choice for calibrating
the 0.5--2.0 keV band and other sources should be used (e.g. white dwarfs and isolated neutron stars).

\section{Fluxes}

\begin{figure*}
\includegraphics[]{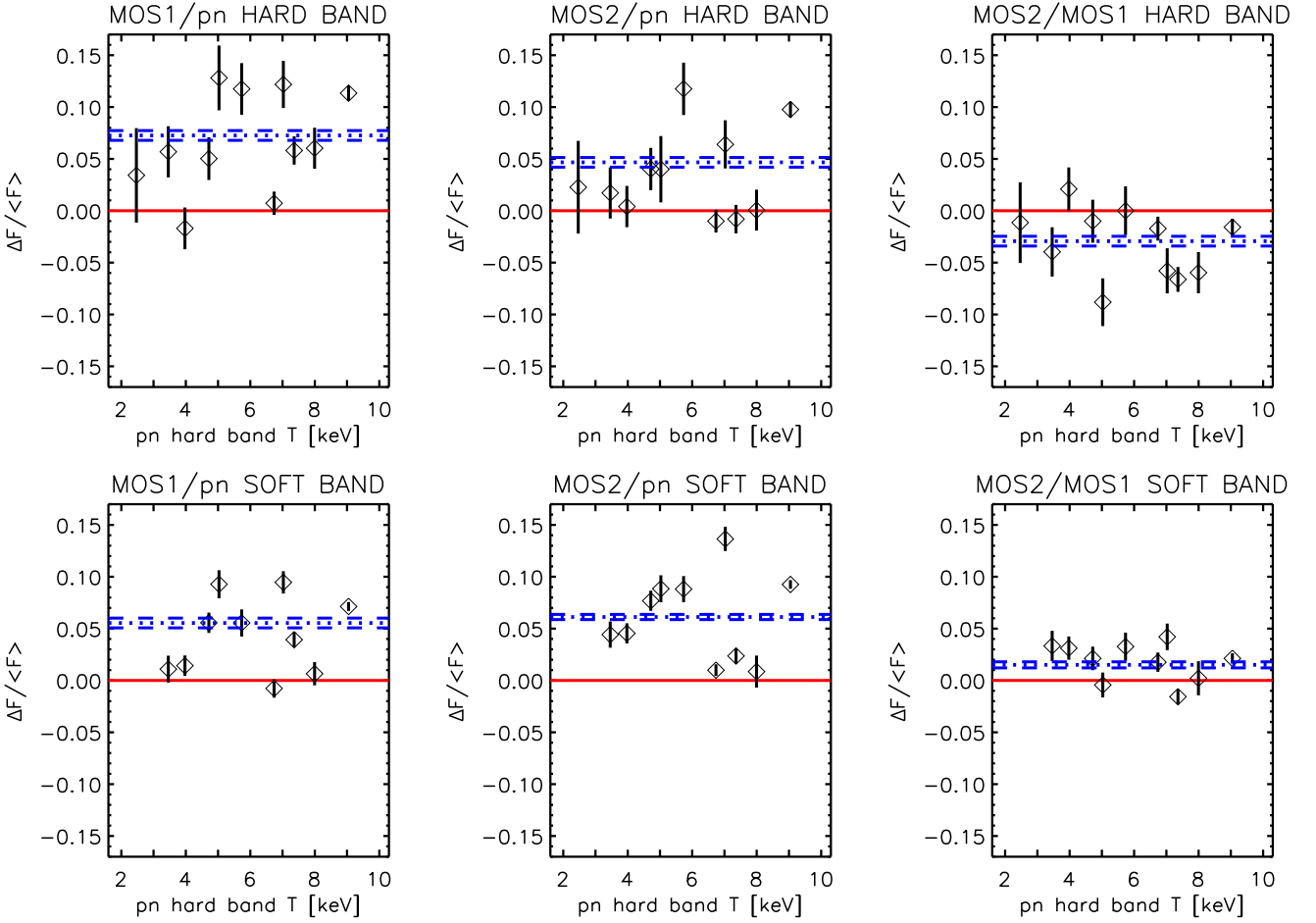}
\caption{The relative flux difference, $f_{F}$ = $\Delta$F/$<$F$>$ (diamonds) and its statistical uncertainties in the 
hard band (upper row) and in the soft band (lower row) are shown as a function of pn hard band temperature for different 
instrument pairs. 
The dotted and dashed lines show the weighted mean of $f_{F}$ $\pm$ the error of the mean. 
\label{pn_mos_flux_plot.fig}}
\end{figure*}

We then examined the fluxes given by the best-fit MEKAL models in the hard and soft bands. We calculated the statistical 
1$\sigma$ uncertainties of the flux measurements by considering the variation of all the free MEKAL parameters.
We limited the pn--MOS comparison within the smaller outer extraction radii used for ACIS (see Sect. \ref{regions}).
Due to the CCD gaps and bad columns in the central regions the fraction of the useful detector area to the full annulus 
given by the extraction radii is less than 100\% (except for ACIS-S), and varies between the instruments for a given 
cluster. Also the exclusion of the point source regions reduces the useful detector area. To enable a meaningful comparison of the fluxes derived with different instruments, we divided the measured flux
by the fraction of the covered full annulus to recover the intrinsic cluster flux measured by each instrument. Since the 
emission is not constant with radius, the linear scaling may produce some differences for the fluxes derived with 
different instruments for a given cluster due to the differences in the covered areas (see below).

Since the half-width of the extraction region is in most cases comparable to the 90\% encircled radius 
($\sim$40--50 arcsec) of EPIC we expect that $\sim$10\% of the flux originating within the extraction region will end up
outside the studied region ($\equiv$ out-flux). We confirmed this by convolving the surface brightness models with the 
pn PSF and calculating the number of counts in different regions. However, the flux originating outside of the extraction 
region and ending up to the extraction region ($\equiv$ in-flux) compensates very accurately the out-flux. Thus the net 
effect is that the true cluster flux within our extraction regions is affected by less than 0.5\% due to PSF scatter in 
our sample. The very accurace cancellation of the in-flux and out-flux is a consequence of an approximately constant 
count rate within annuli of constant width towards smaller or bigger radii within the PSF radius of ~1 arcmin. This in 
turn is due to the opposite effects of decreasing surface brightness and increasing area of the annuli with an increasing
radius.

\begin{table}
\caption{\label{flux.tab}The fluxes in the hard band (2--7 keV) given by the best-fit single-temperature MEKAL model.}
\centering
\scriptsize{
\begin{tabular}{lcccccc}
\hline\hline

        &          &          & pn         & MOS1      & MOS2       & ACIS        \\             
name    & r$_\mathrm{in}$ & r$_\mathrm{out}$ & flux\tablefootmark{a}       & flux\tablefootmark{a}      & flux\tablefootmark{a}       & flux\tablefootmark{a}        \\
        & [']      & [']      &            &           &            &             \\
\hline
A1795   & 1.5      & 2.7 & 1.08$\pm$0.01 & 1.08$\pm$0.01 & 1.07$\pm$0.01 & 1.16$\pm$0.02 \\
A2029   & 1.5      & 2.5 & 1.14$\pm$0.02 & 1.21$\pm$0.02 & 1.14$\pm$0.02 & 1.27$\pm$0.01 \\
A2052   & 1.7      & 2.5 & 0.31$\pm$0.01 & 0.33$\pm$0.01 & 0.31$\pm$0.01 & 0.34$\pm$0.01 \\
A2199   & 2.0      & 2.9 & 0.86$\pm$0.01 & 0.90$\pm$0.01 & 0.89$\pm$0.01 & 0.89$\pm$0.01 \\
A262    & 1.6      & 2.7 & 0.21$\pm$0.01 & 0.22$\pm$0.01 & 0.21$\pm$0.01 & 0.21$\pm$0.01 \\
A3112   & 1.5      & 2.9 & 0.43$\pm$0.01 & 0.49$\pm$0.01 & 0.49$\pm$0.01 & 0.53$\pm$0.01 \\
A3571   & 0.0      & 2.1 & 1.95$\pm$0.02 & 2.07$\pm$0.02 & 1.93$\pm$0.02 & 2.27$\pm$0.03 \\
A85     & 1.5      & 3.0 & 1.10$\pm$0.02 & 1.25$\pm$0.02 & 1.18$\pm$0.02 & 1.27$\pm$0.02 \\
Coma    & 1.0      & 5.0 & 3.83$\pm$0.02 & 4.29$\pm$0.02 & 4.23$\pm$0.02 & 4.34$\pm$0.03 \\
HydraA  & 1.5      & 2.7 & 0.52$\pm$0.01 & 0.59$\pm$0.01 & 0.54$\pm$0.01 & 0.51$\pm$0.01 \\
MKW3S   & 1.5      & 2.5 & 0.38$\pm$0.01 & 0.37$\pm$0.10 & 0.38$\pm$0.10 & 0.42$\pm$0.01 \\
\hline
\end{tabular}}
\tablefoot{
\tablefoottext{a}{The flux is given in units of $10^{-11}$ erg s$^{-1}$ cm$^{-2}$.}
The fluxes are scaled to correspond to the full annulus given by the inner and outer radius  r$_{\rm in}$ and  r$_{\rm out}$. 
}
\end{table}

\subsection{pn v.s. MOS}
We found that while the MOS2 hard band fluxes are in average only $\sim$3\% lower than those of MOS1, the difference is very
significant (6.2$\sigma$, see Fig. \ref{pn_mos_flux_plot.fig} and Table \ref{flux.tab}). There are bigger differences between the pn and the MOS: 
MOS1 hard band fluxes are higher by $\sim$7\% (15.4$\sigma$) while MOS2 hard band fluxes are higher by 5$\sim$\% (10.0$\sigma$)
than those measured by the pn. The differences between the pn and MOS are similar in the soft band.
This flux difference between the pn and MOS was reported earlier by e.g. Stuhlinger et al. (2008)
 and Mateos et al.
(2009). While qualitatively consistent with the analysis of the 2XMM catalogue in Mateos et al. (2009) (when considering
similar off-axis angles (0--2 arcmin) and energy bands (0.5--12 keV)) our pn/MOS differences are somewhat smaller.
Our results are consistent with other IACHEC work based on a sample of blazars (Smith et al., 2010, in prep.)

\subsection{EPIC v.s. ACIS}
We then compared the EPIC fluxes to those obtained using ACIS data and CALDB version 4.2. We found that the scatter of 
the flux between the pn and ACIS is larger ($\sim$6\% ) around the mean value in both bands is much larger than the
statistical uncertainties of the flux measurements ($\sim$1\% ). This implies a systematic uncertainty component
which varies from cluster-to-cluster.
We examined this by restricting the spectral extraction regions to the common sky area where all the cameras have full 
coverage but this did not decrease the scatter significantly. We also experimented using two-temperature models 
with similar results. 
We further examined whether the differences in the shapes of the best-fit models between different instruments 
contribute to the flux scatter. We fitted the MOS and ACIS data with the best-fit pn model, allowing only the 
normalisation to be a free parameter. This did not reduce the scatter either.

\begin{figure}
\includegraphics[width=10cm]{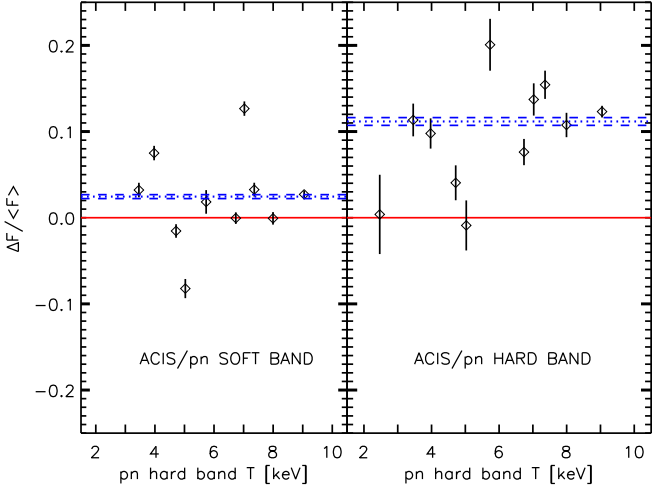}
\caption{The relative flux difference, $f_{F}$ = $\Delta$F/$<$F$>$ (diamonds) and its statistical uncertainties
between ACIS and pn in the soft band (left panel) and the hard band 
(right panel) are plotted against the pn hard band temperature. 
The dotted and dashed lines show the weighted mean of $f_{F}$ $\pm$ the error of the mean. 
\label{acis_pn_flux_plot.fig}}
\end{figure}

In the hard band, the ACIS fluxes are systematically and significantly higher than the pn fluxes, by 11$\pm$0.5\% 
(24.7$\sigma$) on average
(see Fig. \ref{acis_pn_flux_plot.fig}). There is a trend of increasing relative flux difference with increasing temperature. 
ACIS flux values also exceed those of MOS1 and MOS2, but by a smaller amount: 3$\pm$0.5\% (6.1$\sigma$) and 
6$\pm$0.5\% (12.8$\sigma$), respectively.
Thus, our analysis shows that the hard band fluxes using the pn, MOS or ACIS 
instruments, might differ by 5--10\% due to similar uncertainty of the calibration of the normalisation of
the effective areas in the hard band. 

In the soft band, the agreement between ACIS and the pn is much better than in the hard band: the fluxes differ only by 
$\sim$2\% on average (see Fig. \ref{acis_pn_flux_plot.fig}). 
This is interesting when noting that the soft band temperatures between EPIC and ACIS disagreed substantially (see 
Sect. \ref{soft}).

\section{Chandra calibration changes}
\label{caldb}

\begin{figure}
\includegraphics[width=10cm]{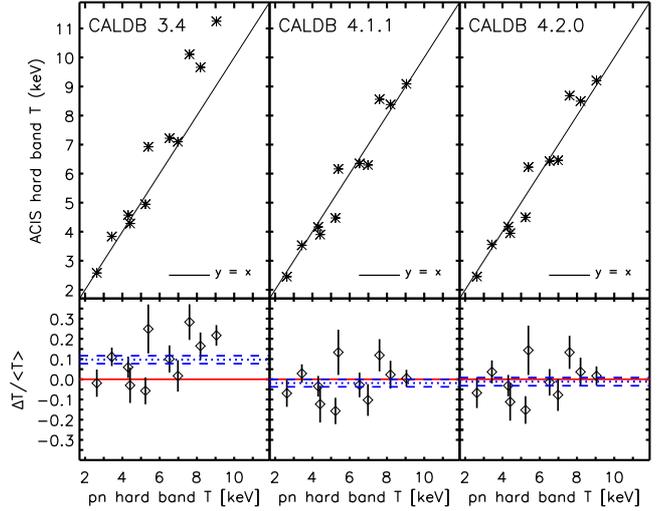}
\caption{The upper panels show the best-fit temperatures (asterisks) in the hard band obtained with \xmm pn 
v.s. those obtained with \chandra ACIS using CALDB 3.4 (left panel), CALDB 4.1.1 (middle panel) and CALDB 4.2 (right 
panel). The lower panels show the relative ACIS-pn temperature difference, $f_{T}$ = $\Delta$T/$<$T$>$ (diamonds) 
and its 1$\sigma$ uncertainty. The dotted and dashed lines show the weighted mean of $f_{F}$ $\pm$ the error of the mean. 
\label{pn_acis_hard.fig}}
\end{figure}

In last few years (2007-2009) a considerable effort has been devoted into re-analysing the ground-based
calibration of \chandra.
Here we examined how the temperatures and fluxes depend on the version of the \chandra calibration data base
(CALDB), i.e. CALDB 3.4 (Dec 2006), CALDB 4.1.1 (Jan 2009) and CALDB 4.2 (Dec 2009), 
(see e.g. David et al., 2007; 
David et al., 2009).

\subsubsection{CALDB 3.4}
After the \chandra launch, gratings observations of AGN showed that the mirror reflectivity near the Ir-M edge was 
15\% higher than predicted by the HRMA effective area model. In CALDB 3.4, a new HRMA effective area model was 
released which included a 22 \AA\ layer of hydrocarbon contaminant on the mirrors and reproduced the observed 
reflectivity across the Ir-M edge. In our earlier work 
(Nevalainen et al. 2007) we found that this model yielded higher hard band temperatures for clusters hotter than 
$\sim$4 keV when compared to values derived with the \xmm EPIC detectors using the calibration as of  Aug 2007. 
Similar results have been obtained with an older version of \chandra calibration (Vikhlinin et al., 2005; 
Kotov et al., 2005; Snowden et al., 2008). 

\begin{figure*}
\includegraphics[]{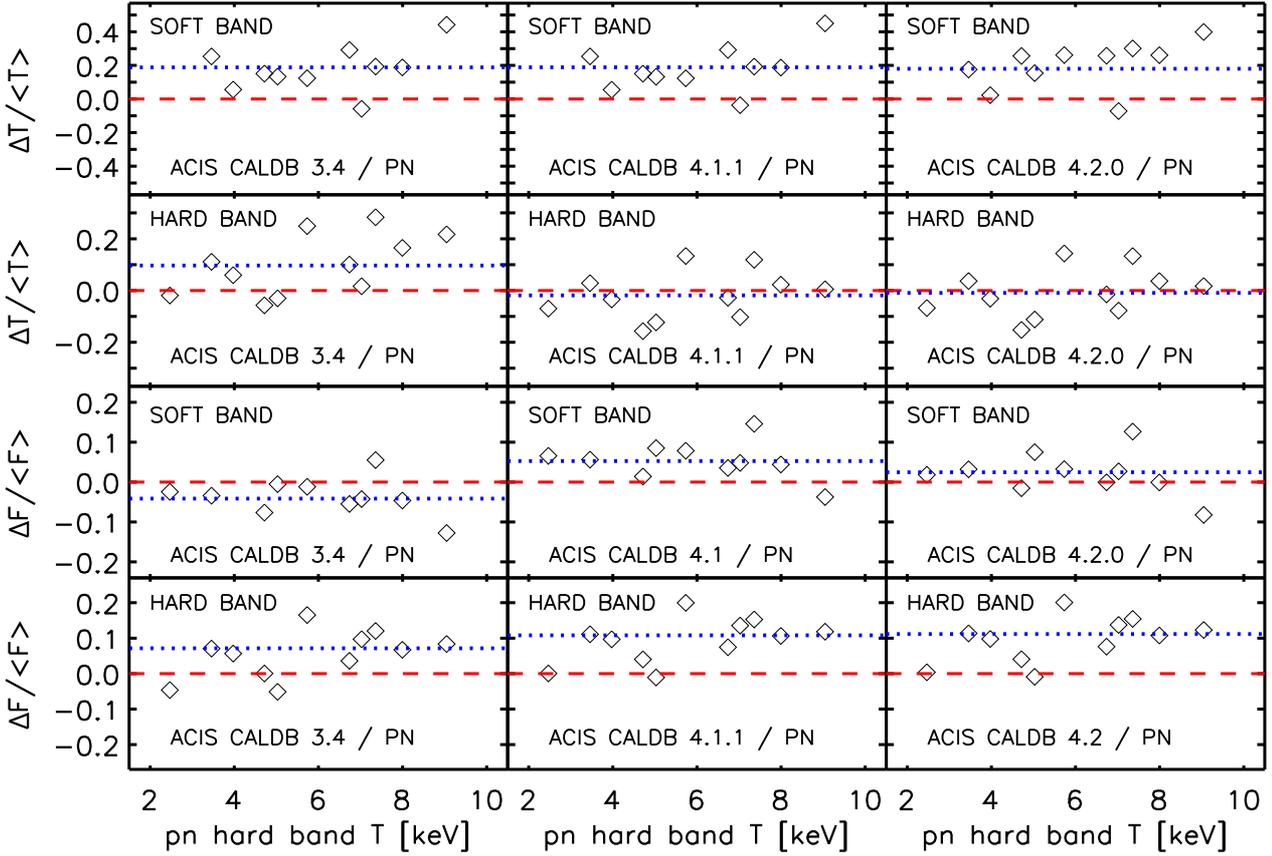}
\caption{
The relative differences of the soft and hard band temperatures (two upper rows) and fluxes (two lower rows) 
between the pn and ACIS at different stages of ACIS calibration (different columns). The values are shown as a function 
of the hard band pn temperatures. Note that the scaling of the y-axis is different in different rows. The average values
are marked with a dotted line.
\label{acis_pn_flux_t_caldb_plot.fig}}
\end{figure*}

In the current work we found that the \chandra CALDB 3.4 hard band temperatures exceeded those obtained with the 
pn using the latest EPIC calibration information in Dec 2009 by an average of $\sim$10\% (see Figs. \ref{pn_acis_hard.fig} and 
\ref{acis_pn_flux_t_caldb_plot.fig} and Table \ref{pn_acis.tab}).  The maximum temperature difference between
the EPIC Dec 2009 calibration and ACIS CALDB 3.4 is $\sim$30\%.

\subsubsection{CALDB 4.1.1}
Following our cluster temperature \chandra/\xmm cross-calibration work (Nevalainen et al. 2007), the ground-based 
HRMA calibration data taken at the X-ray Calibration Facility (XRCF) at MSFC was re-analysed. It was found that the 
molecular contamination was already present on the mirrors at XRCF and that the correction applied to the CALDB3.4 
version of the HRMA effective area over-corrected the effect of the contamination. This problem was corrected in 
CALDB 4.1.1. During XRCF testing, shutters were used to calibrate each of the four mirror shells independently. 
CALDB 4.1.1 contains a model with the XRCF measured depths for the contaminant on each shell.

As a consequence, the change in the reflectivity above 2 keV produced lower hard band temperatures.
We found in the current work that compared to \xmm pn values, the ACIS temperatures obtained with the CALDB 4.1.1 
agreed within a few per cent (see Figs.  \ref{pn_acis_hard.fig} and \ref{acis_pn_flux_t_caldb_plot.fig}). 
Also,  the soft and hard band fluxes increased by $\sim$9\% and $\sim$4\%, respectively.

\subsubsection{CALDB 4.2.0}
ACIS observations of astronomical sources as well as its own external calibration source (ECS) have shown that molecular 
contamination has been building up on the optical/UV blocking filters since launch. A new version (N0005) of the ACIS 
contamination model was released in CALDB 4.2.0.  The previous version of the ACIS contamination model provided a good fit 
to the ECS data  up until about 2006.  For more recent observations, the previous version underestimated the optical depth 
of the contaminant on the ACIS filters. The new version released in CALDB 4.2 produces a much better agreement with the ECS 
data for recent observations.

We found that this revision mostly affected the cluster soft band fluxes, and produced consistency with the pn,
while the soft band temperatures did not change appreciably (see Fig. \ref{acis_pn_flux_t_caldb_plot.fig})


\begin{figure*}
\includegraphics[]{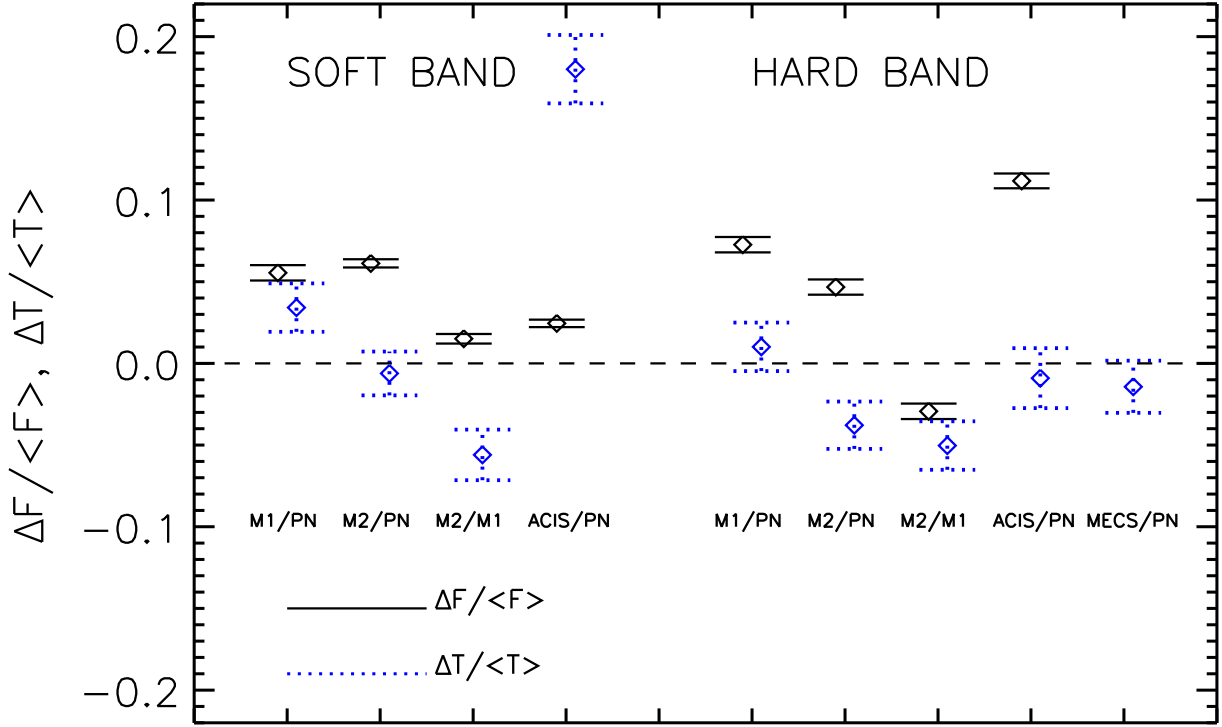}
\vspace{-1.5cm}
\caption{
The average relative difference (diamonds) $\pm$ the error of the mean of the fluxes (solid line) and 
temperatures (dotted line) for different instrument pairs
in the soft band (left side of the plot) and in the 
hard band (right side of the plot).
\label{pn_mos_acis_flux_t_plot.fig}}
\end{figure*}

\begin{table*}
\caption{\label{statres.tab} The relative difference and its significance of temperatures and fluxes 
for different combinations of instruments and methods.} 
\centering
{\normalsize
\begin{tabular}{l|cc|cc|cc|cc|cc|cc|cc|cc}
\hline\hline
Instruments & \multicolumn{2}{c}{T$_\mathrm{hard}$\tablefootmark{a}} & \multicolumn{2}{c}{flux$_\mathrm{hard}$\tablefootmark{a}} &\multicolumn{2}{c}{T$_\mathrm{soft}$\tablefootmark{b}} & 
\multicolumn{2}{c}{flux$_\mathrm{soft}$\tablefootmark{b}} &  \multicolumn{2}{c}{T$_\mathrm{soft}$/T$_\mathrm{hard}$} &  \multicolumn{2}{c}{T$_\mathrm{wide}$\tablefootmark{c}} &
\multicolumn{2}{c}{T$_\mathrm{Fe}$\tablefootmark{d}} &
\multicolumn{2}{c}{T$_\mathrm{cont}$\tablefootmark{e} / T$_\mathrm{Fe}$}  
\\
          & $\mu$\tablefootmark{e} & sig.\tablefootmark{f}  & $\mu$\tablefootmark{e} & sig.\tablefootmark{f}  & 
$\mu$\tablefootmark{e} & sig.\tablefootmark{f} & $\mu$\tablefootmark{e} & sig.\tablefootmark{f} & 
$\mu$\tablefootmark{e} & sig.\tablefootmark{f} & $\mu$\tablefootmark{e} & sig.\tablefootmark{f} & 
$\mu$\tablefootmark{e} & sig.\tablefootmark{f} & $\mu$\tablefootmark{e} & sig.\tablefootmark{f} \\ 
\hline
MOS2/MOS1 & -5 & 3.4 & -3 & 6.2 & -6 & 3.6 & 2 & 5.2 & $\ldots $ & $\ldots $ & -5 & 9.9  &   $\ldots $     &   $\ldots $     &  $\ldots $     &   $\ldots $     \\ 
MOS1/pn   &  1 & 0.7 & 7 & 15.4 &  5 & 3.7 & 6 & 11.8 & $\ldots $ & $\ldots $ &  5 & 10.3 &  $\ldots $      &  $\ldots $      &  $\ldots $     &   $\ldots $     \\
MOS2/pn   & -4 & 2.6 & 5 & 10.0 & -1 & 0.5 & 6 & 24.3 & $\ldots $ & $\ldots $ &  0 & 0.2  &  $\ldots $      &  $\ldots $      &  $\ldots $     &   $\ldots $     \\
MECS/pn   & -1 & 0.9 & $\ldots$      &  $\ldots$       &  $\ldots$     & $\ldots$      & $\ldots$      &  $\ldots$     &  $\ldots$     &  $\ldots $      &   $\ldots $    &  $\ldots $    &  $\ldots $      &   $\ldots $     &  $\ldots $     &   $\ldots $     \\
ACIS/pn   & -1 & 0.6 & 11 & 24.7 & 18 & 8.6 & 2 & 10.6 & $\ldots $ & $\ldots $ & 14 & 18.0 &  $\ldots $      &   $\ldots $     &  $\ldots $     &   $\ldots $     \\
ACIS/MOS1 &  $\ldots$     &  $\ldots$       & 3  & 6.1    & $\ldots$ & $\ldots$ & -3 & 9.8 &  $\ldots $     &   $\ldots $     & 8     &  10.0    &  $\ldots $      &   $\ldots $     &   $\ldots $    &   $\ldots $     \\
ACIS/MOS2 &  $\ldots$     &  $\ldots$       & 6  & 12.8    & $\ldots$ & $\ldots$ & -5 & 16.7 &  $\ldots $     &   $\ldots $   & 15    &  18.9    &   $\ldots $     &   $\ldots $     &   $\ldots $    &   $\ldots $     \\
MOS/pn    &  $\ldots$     &  $\ldots$       &  $\ldots$     &  $\ldots$       &  $\ldots$     & $\ldots$      & $\ldots$      &  $\ldots$    &  $\ldots $     &   $\ldots $     &  $\ldots $     &  $\ldots $    & -6 & 2.2 &   $\ldots $    &   $\ldots $     \\
pn        &  $\ldots$     &  $\ldots$       &  $\ldots$     & $\ldots$        & $\ldots$      & $\ldots$      &  $\ldots$    &   $\ldots$   & -36 & 29.9   &   $\ldots $    &  $\ldots $    &   $\ldots $     &  $\ldots $      & 1 & 0.2 \\
MOS       &  $\ldots$     &  $\ldots$       &  $\ldots$     & $\ldots$        & $\ldots$      &$\ldots$       &  $\ldots$    &  $\ldots$    &  $\ldots$     &  $\ldots $      &  $\ldots $     &  $\ldots $    &  $\ldots $      &  $\ldots $      & 7 & 1.6 \\
MOS1      &  $\ldots$     &  $\ldots$       & $\ldots$      & $\ldots$        & $\ldots$      & $\ldots$      &  $\ldots$    & $\ldots$    & -32 & 21.8  &  $\ldots $     &  $\ldots $    &   $\ldots $     &  $\ldots $      &  $\ldots $     &   $\ldots $     \\
MOS2      &  $\ldots$     &   $\ldots$      & $\ldots$      &$\ldots$         & $\ldots$      & $\ldots$      &  $\ldots$    & $\ldots$    & -32 & 23.1   &  $\ldots $     &  $\ldots $    &  $\ldots $      &  $\ldots $      &  $\ldots $     &   $\ldots $     \\
ACIS      & $\ldots$      & $\ldots$        & $\ldots$      & $\ldots$        &$\ldots$       & $\ldots$      &  $\ldots$    & $\ldots$    & -12 & 1.6  &  $\ldots $     &  $\ldots $    &  $\ldots $      &  $\ldots $      &   $\ldots $    &   $\ldots $     \\
\hline
\end{tabular}}
\tablefoot{
The values correspond to the best-fit single temperature models in the   
\tablefoottext{a}{2--7 keV band}, 
\tablefoottext{b}{0.5--2 keV band}, 
\tablefoottext{c}{0.5--7 keV band}, 
\tablefoottext{d}{$\sim$6.0--7.0 keV band} and 
\tablefoottext{e}{2--6 keV band}
The ACIS values are obtained using CALDB 4.2.0.
\tablefoottext{e}{$\mu$ gives the weighted mean of the relative difference between the two measurements in percentages.}
\tablefoottext{f}{sig. gives the significance of the relative difference in terms of $\sigma$.}
}
\end{table*}

\section{Conclusions}
We performed an X-ray spectral analysis of clusters of galaxies using data obtained with the \xmm EPIC instruments pn, 
MOS1 and MOS2 and \chandra ACIS-S and ACIS-I. We additionally used the published results obtained using \sax MECS 
(deGrandi \& Molendi, 2002). We compared the derived temperatures and fluxes for each cluster based on the 
current instrument calibrations as of December 2009. The results are summarised in Fig. \ref{pn_mos_acis_flux_t_plot.fig}
and Table \ref{statres.tab}.

We found that there are no systematic differences in the temperatures obtained by fitting the 2--7 keV energy band of ACIS
and EPIC (2--10 keV for MECS). These values are also consistent with those obtained by the EPIC temperature measurements based
on the Fe XXV/XXVI line ratio. This shows that the energy dependence of the effective area  in this band is accurately 
modelled in the studied instruments.  Thus, the IACHEC cluster sample in the studied radial range 
($\sim$ 0.1--0.3 r$_{500}$) constitutes a set of standard candles for the calibration of the energy dependence of the 
hard band effective area. On the other hand, the disagreements at 6--25$\sigma$ level on the hard band fluxes showed that 
there are systematic calibration uncertainties in the normalisations of the effective areas by 5--10\% in the 2--7 keV band. 

The temperatures obtained by fitting the soft band (0.5--2.0 keV) data of the pn and ACIS differ systematically and 
significantly, by $\sim$18\% (i.e. at 8.6$\sigma$ level) on average. This indicates remaining uncertainties with the 
calibration of the energy dependence of the effective area in the 0.5--2.0 keV band in one or all instruments. 
Comparison of the residuals showed that the relative pn/ACIS cross-calibration bias is approximately a linear function of 
energy in the 0.5--2.0 keV band, amounting to a variation of 10\% in this band. The uncertainties of the 
calibration and the modelling of the cluster emission in the soft band cannot be simultaneously resolved by using 
clusters of galaxies alone.

Due to the higher effective area and the higher number of intrinsic cluster photons in the soft band, the statistical 
weight of the soft band data is much higher than that of the hard band. Thus, the calibration uncertainties in the soft 
band will affect the scientific analysis of clusters of galaxies, when using the full useful energy band ($\sim $0.5--7.0 
keV). Considering the variation between the different instruments, the 0.5--7.0 keV band temperature measurement of 
clusters of galaxies with EPIC/\xmm or ACIS/\chandra is uncertain by 10--15\% on average. 
These uncertainties will also affect the analysis of the wide band continuum spectra of other types of objects 
using ACIS or EPIC.

We evaluated the systematic effects on the Fe XXV/XXVI line ratio temperature measurement due to uncertainties of 
the implemented EPIC calibration of the energy scale and energy resolution and redistribution and the details of the 
line ratio modeling. The effect on the measured temperature is $\sim$4\%. The temperatures measured using the continuum 
shape and the Fe XXV/XXVI line ratio agree very accurately. This indicates that the deviations from the ionisation 
equilibrium state and Maxwellian electron velocity distribution are negligible in the studied regions of this cluster 
sample. Since the Fe XXV/XXVI line ratio measurement is not sensitive to smooth changes in the calibration of the 
effective area, it could be a powerful hard band calibration tool in future X-ray missions.

After the submission of this paper an update on the pn redistribution become public (Haberl et al., 2010). 
This may improve the Fe XXV/XXVI emission line modeling and we will address this issue in a follow-up paper. The effect 
on the fluxes and continuum temperatures is likely small, because the line emission (besides Fe XXV and XXVI) is weak in 
the cluster sample. A study of a \xmm sample of sources with a wide range of  spectral shapes (Stuhlinger et al. 
2010, in prep.) shows that the change of the relative pn/MOS flux above 0.5 keV between the new calibration and that used
in the present paper is 3\% at the most. Thus, the cluster flux discrepancy between the pn and MOS will not be resolved 
with the new pn calibration in July 2010.

\begin{acknowledgements}
The work is based on observations obtained with \xmm, an ESA science mission with instruments and contributions 
directly funded by ESA Member States and NASA. This research has made use of data and software provided by the 
\chandra X-ray Center (CXC).
JN is supported by the Academy of Finland. We thank M. Bonamente, S. Molendi, P. Plucinsky and M. Smith for help and 
useful comments. We thank the IACHEC team for support. 
\end{acknowledgements}

\clearpage

\begin{appendix}

\section{Spectral fits}
\label{app_plots}
\begin{figure*}
\includegraphics[width=23cm,angle=-90]{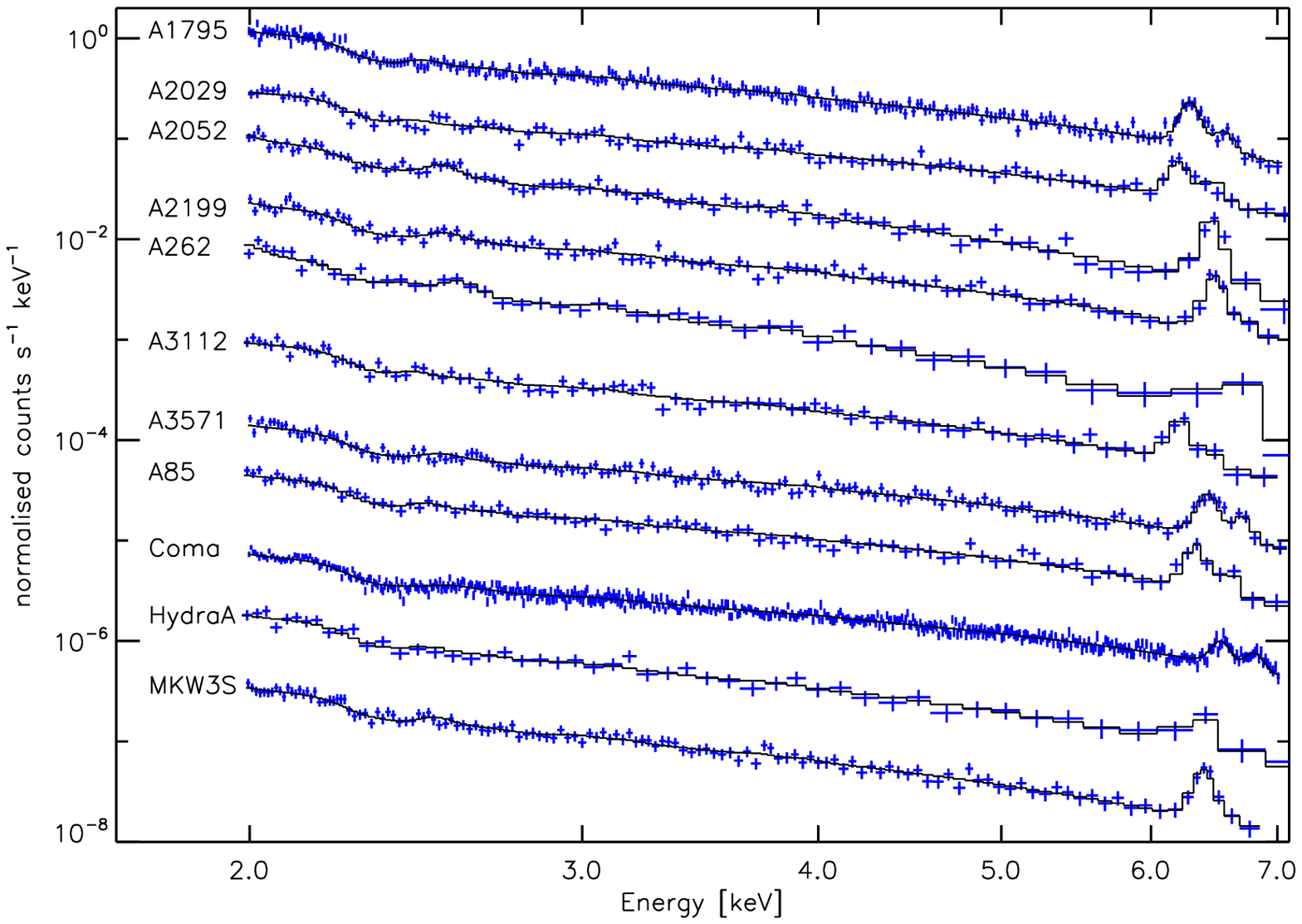}
\caption{The hard band pn spectra (crosses) and the best-fit single-temperature fits (solid lines) for the cluster 
sample. The normalisation of the spectra is adjusted for plot clarity and does not 
correctly reflect the relative brightness of the clusters. 
\label{pn_hardplot2.fig}}
\end{figure*}

\begin{figure*}
\includegraphics[width=23cm,angle=-90]{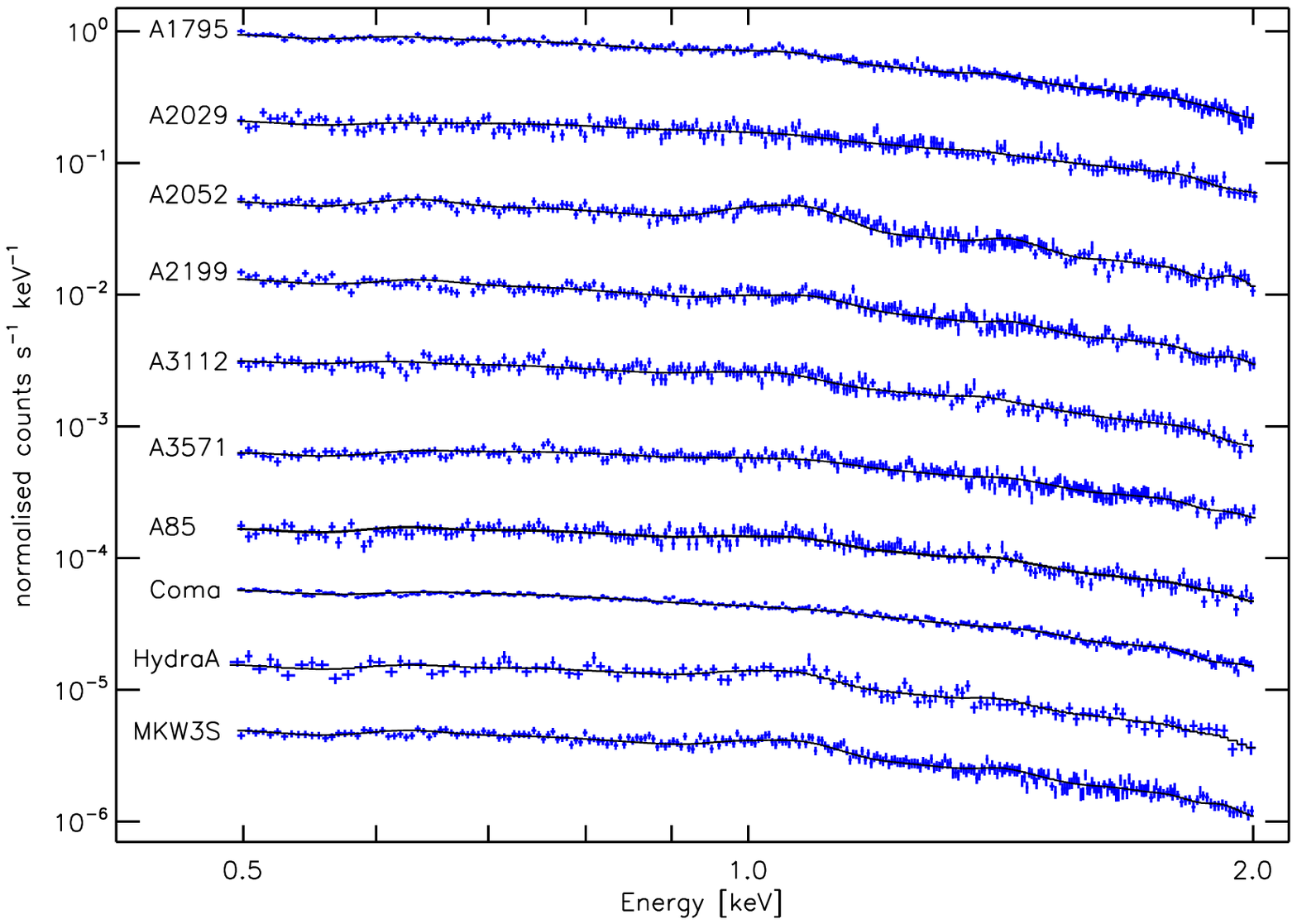}
\caption{The soft band pn spectra (crosses) and the best-fit single-temperature fits (solid lines) for the cluster 
sample. The normalisation of the spectra is adjusted for plot clarity and does not 
correctly reflect the relative brightness of the clusters. 
\label{pn_softplot2.fig}}
\end{figure*}

\begin{figure*}
\includegraphics[width=23cm,angle=-90]{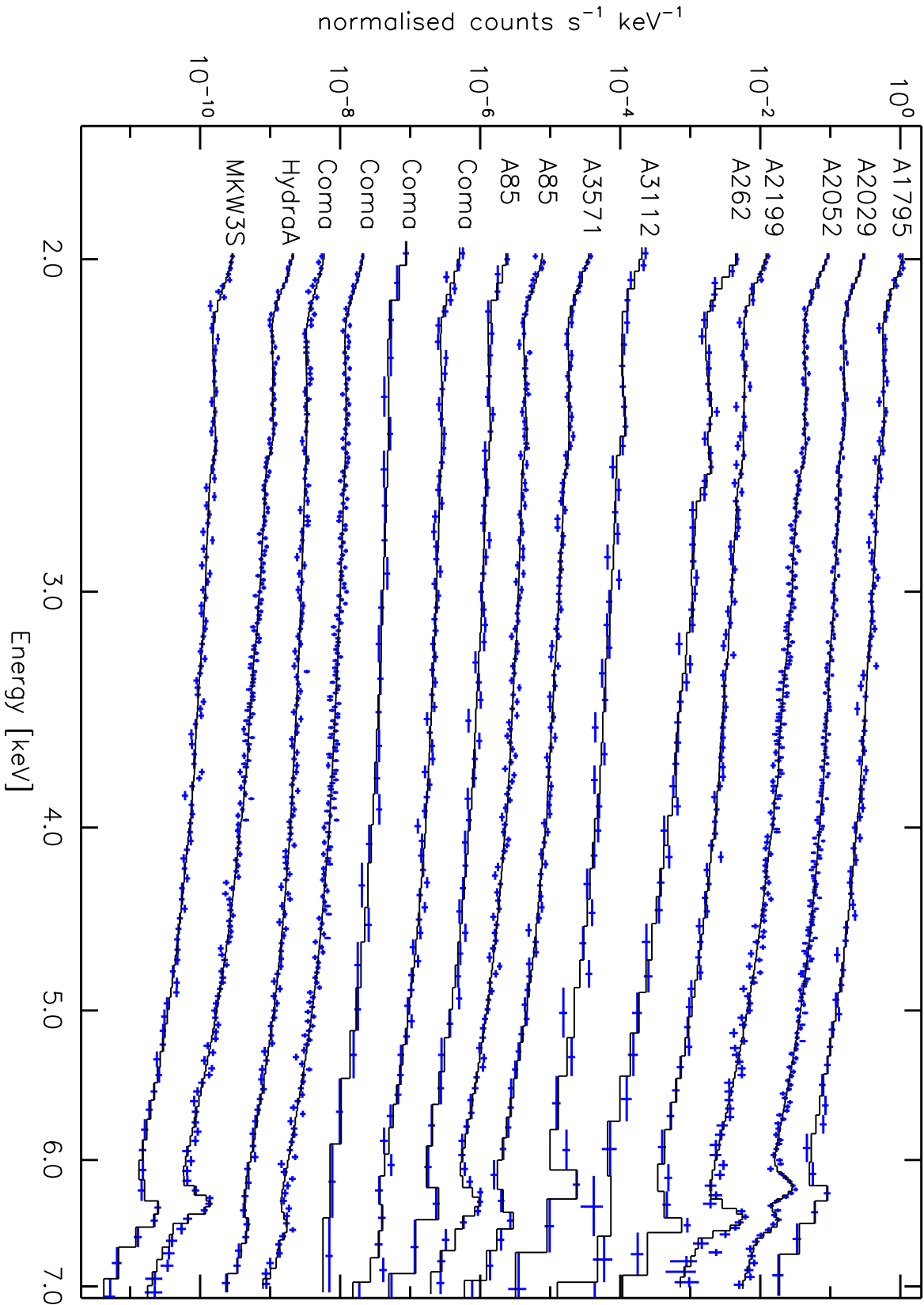}
\caption{The hard band ACIS spectral data (crosses) and the best-fit single-temperature fits (lines) for the cluster 
sample. The normalisation of the spectra is adjusted for plot clarity and does not 
reflect correctly the relative brightness of the clusters. The spectra for A85 and Coma were obtained from several 
CCD chips and were fitted simultaneously. 
\label{acis_hardplot2.fig}}
\end{figure*}

\begin{figure*}
\includegraphics[width=23cm,angle=-90]{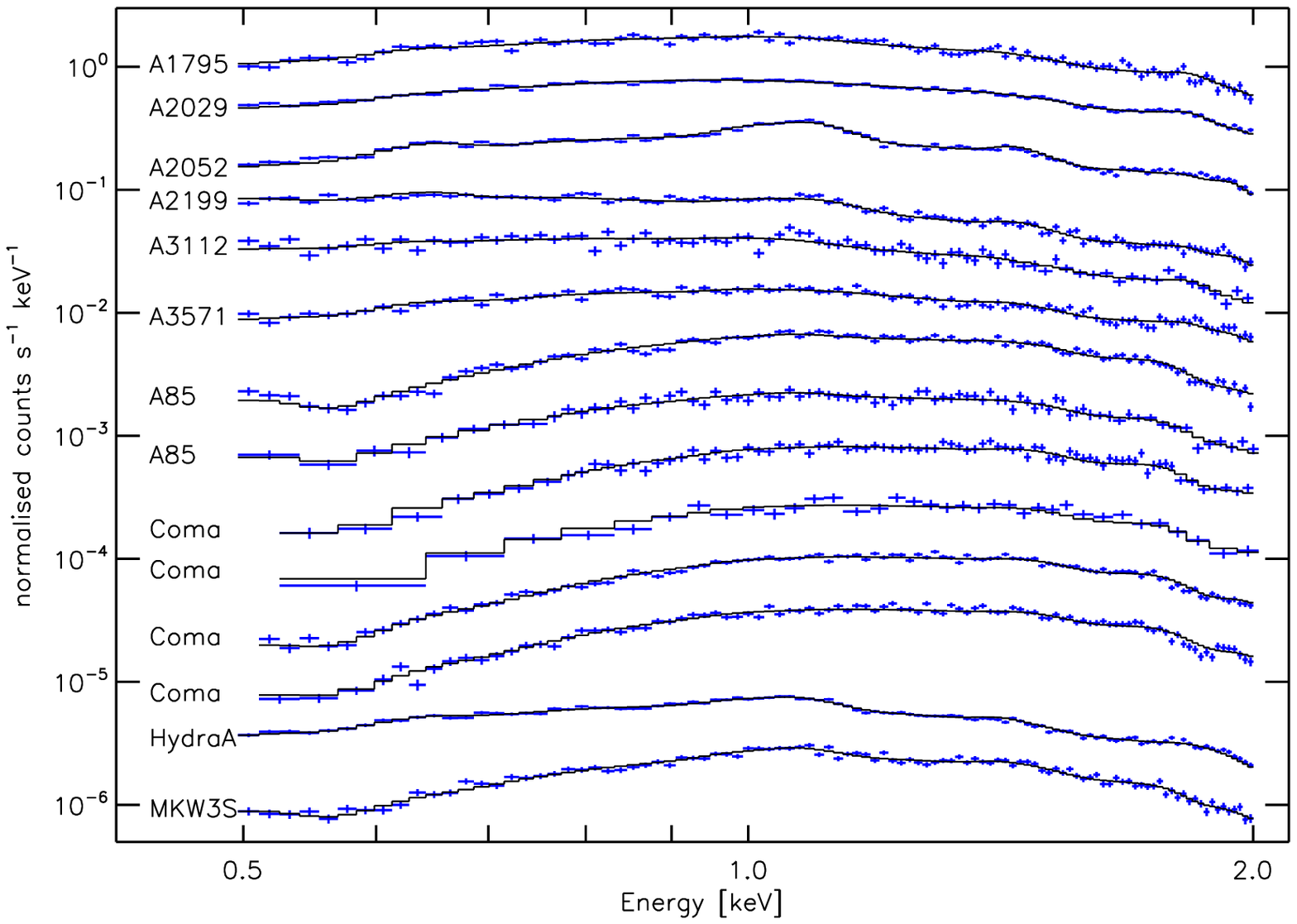}
\caption{The soft band ACIS spectra (crosses) and the best-fit single-temperature fits (solid lines) for the cluster 
sample. The normalisation of the spectra is adjusted for plot clarity and does not 
correctly reflect the relative brightness of the clusters. The spectra for A85 and Coma were obtained from several 
CCD chips and were fitted simultaneously. 
\label{acis_softplot2.fig}}
\end{figure*}

We show here the data and the best-fit single-temperature MEKAL models in the hard and soft bands for 
pn (Figs. \ref{pn_hardplot2.fig} and \ref{pn_softplot2.fig}) and for ACIS
(Figs. \ref{acis_hardplot2.fig} and \ref{acis_softplot2.fig}) used for the pn/ACIS temperature and flux comparison.
The spectral parameters of the pn and ACIS fits in all bands are shown in Table 
\ref{pn_acis.tab}.

\begin{table*}[h]
\caption{\label{pn_acis.tab}Best-fit temperatures and metal abundances for pn and ACIS in different bands}
\centering
{\normalsize
\begin{tabular}{lcc|ccc|ccc}
\hline\hline
        &          &          & \multicolumn{3}{c}{pn} & \multicolumn{3}{c}{ACIS CALDB 4.2.0}  \\ 
name    & r$_{\rm in}$ & r$_{\rm out}$ &  T & abund & $\chi^2$/dof & T & abund & $\chi^2$/dof \\
        & [']     & [']       & [keV] & [Solar] & & [keV] & [Solar] & \\ 
\hline
\multicolumn{9}{c}{ } \\
\multicolumn{9}{c}{\bf HARD BAND (2.0--7.0 keV)} \\
\hline
A1795   & 1.5 & 2.7 & 6.5[6.3--6.7] & 0.48[0.44--0.51] & 204.7/280 & 6.4[6.1--6.8]  & 0.45[0.38--0.52] & 92.8/86   \\
A2029   & 1.5 & 2.5 & 8.2[7.7--8.7] & 0.53[0.45--0.60] & 79.0/112  & 8.5[8.2--8.8]  & 0.52[0.48--0.55] & 265.1/258 \\
A2052   & 1.7 & 2.5 & 3.4[3.3--3.6] & 0.56[0.49--0.63] & 59.9/91   & 3.6[3.4--3.7]  & 0.44[0.40--0.48] & 193.2/193 \\
A2199   & 2.0 & 2.9 & 5.2[5.0--5.5] & 0.52[0.46--0.58] & 80.0/111  & 4.5[4.3--4.7]  & 0.44[0.37--0.52] & 89.6/79   \\
A262    & 1.6 & 2.7 & 2.6[2.5--2.8] & 0.44[0.32--0.57] & 28.5/46   & 2.5[2.3--2.6]  & 0.62[0.50--0.75] & 50.3/45   \\
A3112   & 1.5 & 2.9 & 5.4[5.1--5.7] & 0.37[0.31--0.44] & 56.8/88   & 6.2[5.6--6.9]  & 0.93[0.77--1.10] & 66.5/36 \\
A3571   & 0.0 & 2.1 & 7.6[7.3--8.0] & 0.59[0.54--0.65] & 119.1/166 & 8.7[8.1--9.3]  & 0.71[0.61--0.83] & 65.8/89   \\
A85     & 1.5 & 3.0 & 7.0[6.5--7.5] & 0.56[0.48--0.64] & 47.1/95   & 6.5[6.2--6.7]  & 0.52[0.47--0.57] & 190.3/165 \\
Coma    & 1.0 & 5.0 & 9.0[8.8--9.3] & 0.33[0.31--0.35] & 475.5/557 & 9.2[8.9--9.6]  & 0.40[0.36--0.44] & 457.8/438 \\
HydraA  & 1.5 & 2.7 & 4.4[4.1--4.8] & 0.25[0.14--0.35] & 25.4/48   & 4.0[3.8--4.1]  & 0.35[0.31--0.38] & 179.4/187 \\
MKW3S   & 1.5 & 2.5 & 4.3[4.1--4.5] & 0.39[0.34--0.44] & 76.4/134  & 4.2[4.0--4.3]  & 0.45[0.40--0.51] & 101.6/99  \\
\hline
\multicolumn{9}{c}{ } \\
\multicolumn{9}{c}{\bf SOFT BAND (0.5-2.0 keV)} \\
\hline
A1795   & 1.5 & 2.7 & 4.7[4.5--4.8] & 0.33[0.29--0.36] & 232.6/303 & 6.0[5.6--6.5] & 0.34[0.24--0.45] & 115.6/100 \\ 
A2029   & 1.5 & 2.5 & 6.1[5.7--6.6] & 0.30[0.19--0.41] & 203.1/231 & 8.0[7.7--8.3] & 0.31[0.23--0.40] & 115.7/100 \\ 
A2052   & 1.7 & 2.5 & 2.7[2.6--2.8] & 0.48[0.45--0.52] & 226.7/250 & 3.2[3.2--3.3] & 0.61[0.58--0.64] & 165.8/100 \\ 
A2199   & 2.0 & 2.9 & 3.8[3.7--4.0] & 0.40[0.35--0.45] & 251.8/255 & 5.0[4.7--5.2] & 0.70[0.63--0.80] & 104.9/100 \\  
A3112   & 1.5 & 2.9 & 3.9[3.7--4.1] & 0.35[0.29--0.41] & 196.5/218 & 5.1[4.6--5.6] & 0.33[0.21--0.46] & 106.1/82 \\  
A3571   & 0.0 & 2.1 & 5.7[5.3--6.0] & 0.39[0.31--0.47] & 258.3/276 & 7.7[7.0--8.5] & 0.69[0.51--0.91] & 110.2/100 \\  
A85     & 1.5 & 3.0 & 5.2[4.9--5.6] & 0.50[0.41--0.60] & 208.3/220 & 4.9[4.7--5.1] & 0.23[0.17--0.28] & 200.8/180 \\
Coma    & 1.0 & 5.0 & 6.0[5.8--6.1] & 0.19[0.15--0.22] & 325.7/303 & 8.9[8.5--9.4] & 0.63[0.49--0.82] & 368.8/315 \\  
HydraA  & 1.5 & 2.7 & 2.9[2.7--3.0] & 0.28[0.24--0.33] & 106.7/131 & 3.3[3.3--3.4] & 0.37[0.35--0.40] & 96.7/100 \\ 
MKW3S   & 1.5 & 2.5 & 3.2[3.1--3.3] & 0.36[0.33--0.40] & 220.6/277 & 3.3[3.1--3.4] & 0.32[0.28--0.36] & 116.3/100 \\  
\hline
\multicolumn{9}{c}{ } \\
\multicolumn{9}{c}{\bf WIDE BAND (0.5-7.0 keV)} \\
\hline
A1795   & 1.5 & 2.7 & 5.4[5.4--5.5]  & 0.44[0.42--0.47] & 509.5/585 & 6.3[6.2--6.5] & 0.42[0.37--0.49] & 208.4/188 \\ 
A2029   & 1.5 & 2.5 & 7.0[6.8--7.2]  & 0.47[0.42--0.53] & 294.4/345 & 8.9[8.8--9.1] & 0.52[0.48--0.55] & 398.1/360 \\ 
A2052   & 1.7 & 2.5 & 3.0[2.9--3.0]  & 0.56[0.53--0.60] & 308.2/343 & 3.3[3.3--3.4] & 0.60[0.57--0.62] & 387.5/295 \\ 
A2199   & 2.0 & 2.9 & 4.2[4.1--4.3]  & 0.49[0.45--0.53] & 368.4/368 & 4.7[4.6--4.8] & 0.57[0.52--0.62] & 200.4/181 \\ 
A3112   & 1.5 & 2.9 & 4.4[4.3--4.5]  & 0.41[0.36--0.45] & 270.7/308 & 5.7[5.5--5.9] & 0.65[0.55--0.76] & 180.7/120 \\ 
A3571   & 0.0 & 2.1 & 6.6[6.4--6.7]  & 0.53[0.49--0.57] & 395.7/444 & 8.6[8.4--8.8] & 0.72[0.64--0.81] & 176.7/191 \\ 
A85     & 1.5 & 3.0 & 5.5[5.4--5.7]  & 0.53[0.47--0.58] & 272.0/317 & 5.8[5.7--5.9] & 0.46[0.43--0.51] & 415.3/347 \\  
Coma    & 1.0 & 5.0 & 7.1[7.0--7.2]  & 0.28[0.27--0.30] & 969.4/862 & 9.0[8.8--9.1] & 0.40[0.36--0.44] & 830.2/753 \\ 
HydraA  & 1.5 & 2.7 & 3.3[3.2--3.4]  & 0.33[0.28--0.39] & 152.9/181 & 3.8[3.7--3.8] & 0.41[0.39--0.43] & 312.9/289 \\ 
MKW3S   & 1.5 & 2.5 & 3.6[3.5--3.6]  & 0.42[0.41--0.45] & 335.6/413 & 3.8[3.7--3.8] & 0.44[0.41--0.48] & 241.1/201 \\ 
\hline
\end{tabular}}
\tablefoot{
Temperatures and their uncertainties at 1$\sigma$ level were obtained with a single-temperature MEKAL model using data 
from the pn and ACIS instruments in different bands}
\end{table*}

\clearpage

\section{Systematic uncertainties affecting the Fe XXV/XXVI line ratio measurement using \xmm EPIC instruments}
\label{sys_app}

\subsection{Effective area}
\label{sys_tot}
\subsubsection{Break in the effective area}
The virtue of the Fe XXV/XXVI line ratio method is that the energies of the two lines only differ by $\sim$300 eV.
If there was a sudden change in the real effective area, co-incidentally at an energy between those of the FeXXV and 
FeXXVI line energies, that was not correctly implemented into the calibration, then the measured line ratio would 
be inaccurate. However, the redshifts in our sample vary from 0.0231 to 0.0773 which produce a $\sim$300 eV variation in 
the Fe line energies in different clusters. Thus, a possible break in the effective area at a fixed energy cannot produce
a systematic effect in the sample, but can affect some of the clusters. To estimate the effect of this unlikely 
situation, we simulated a pn spectrum of a cluster with T=9 keV and metal abundance of 0.3 Solar at a redshift z=0. When 
fitting the simulated data in a 6.45--7.25 keV band, we introduced a 10\% drop in the effective area in the associated 
auxiliary response file at E$\ge$6.85 keV. In order to produce the data above 6.85 keV, the model Fe XXVI flux increases,
i.e. the temperature increased to 10.4 keV, which is 15\% higher than the input value.

\subsubsection{Smooth calibration bias}
\label{effareasimu}
We note that we found an excellent agreement between the hard band temperatures obtained with different instruments (see 
Sect. \ref{hard}). Because the different instruments are calibrated based on a combination of ground measurements and 
in-flight measurements of different types of celestial sources, it would be a highly unlikely co-incidence 
that all the instruments have similar uncertainties in the calibration as to yield similarly biased temperatures. 
This hypothetical bias should be a smooth function of energy in order for the biased model to agree well with the data, 
as observed. We examined the sensitivity of the Fe XXV/XXVI line ratio temperature measurement to a smooth hard band 
calibration bias using simulated data. We created spectra with kT in the range 4--10 keV. We kept the metal abundance 
at 0.3 Solar and the redshift at 0. We used 10$^{7}$ counts in the 6.45--7.25 keV energy band in order to eliminate the 
statistical effects. We used the auxiliary response file of A1795 for the simulations. 

When fitting the simulated data, we approximated the effect of a smooth calibration bias by modifying the auxiliary file 
we used for the simulations. In detail, we multiplied the effective area column with a linear function which has a value 
of 1.0 at 1 keV, and varies between 0.8 and 1.2 at 10 keV. Thus the effective area is unchanged at 1 keV, but is 
underestimated or overestimated by a maximum of  20\% at 10 keV.

\begin{figure}
\includegraphics[width=9cm,angle=0]{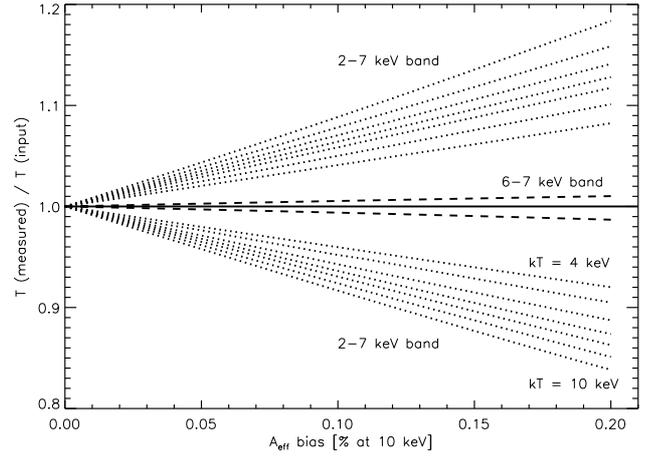}
\caption{The temperature bias in the simulated spectra as a function of the fraction of the linear bias in the 
effective area at 10 keV for the 2.0 -- 7.0 keV band fits (dotted lines) and for the $\sim$6--7 keV band fit (dashed 
lines). The different dotted lines denote different input temperatures from 4 to 10 keV (higher temperatures yield higher
bias). For the $\sim$6--7 keV band fit only the curves for kT=10 keV are shown for clarity. The lines in the upper half 
(lower half) correspond to positive (negative) bias in the effective area. 
\label{simu_arf.fig}}
\end{figure}

We found that the resulting relative bias in the temperature measurement is higher for the higher input temperatures,
due to their higher statistical weight in the more biased high energy band (see Fig.  \ref{simu_arf.fig}). However, the 
effect in the Fe XXV/XXVI temperature measurement was very small: even if the effective area was off by 15\% at photon 
energies of 10 keV, the best-fit Fe XXV/XXVI temperature would be biased by less than 1\%  (see Fig. \ref{simu_arf.fig}). 
The reason for this is that even though the effective area is over- or underpredicted by up to $\sim$15\%  in the 
6--7 keV band in average, the relative change of the effective area between the 6.45 and 7.25 keV is only $\sim$1\%. Thus
the only effect is a biased MEKAL normalisation up to $\pm \sim$10\%, while the shape of the spectrum does not change.

The situation is the opposite in the 2.0--7.0 keV band. The varying degree of calibration bias in a 10--20\% range 
does not much affect the derived emission measures (which only vary by $\sim$1\%). However, it does affect the measured 
continuum shape, yielding a 10--20\% bias in the temperature (see Fig. \ref{simu_arf.fig}).

\subsubsection{ACIS calibration changes}
We further tested the effects of smooth effective area calibration changes on the Fe XXV/XXVI line ratio measurement by a 
comparison of ACIS temperatures derived using calibration versions CALDB 3.4 and CALDB 4.2.0. Note that the substantial, 
smooth changes in the ACIS effective area calibration between these two versions resulted in significant changes in the 
hard and soft band  temperatures (see Sect. \ref{hard} and \ref{soft}). We relaxed the requirement for the minimum 
number of counts to 400 since the bias due to the limited number of counts (see Sect. \ref{fe_stat}) does not 
depend on the calibration version and thus will not introduce a systematic difference. 

\begin{figure*}
\includegraphics[width=15cm]{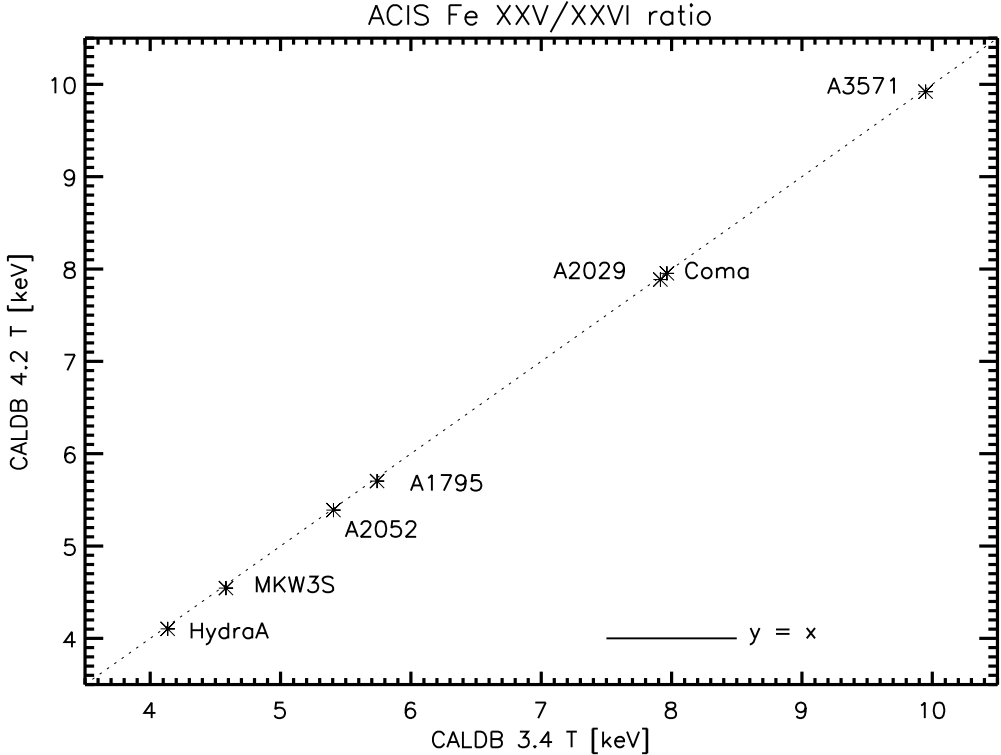}
\caption{The best-fit temperatures obtained from ACIS data in the $\sim$6--7 keV energy band using calibration
versions CALDB 3.4 v.s CALDB4.2.0.
\label{acis_old_new_6070.fig}}
\end{figure*}

We fitted the data in the $\sim$6--7 keV energy band allowing the emission measure to be a free parameter. We found that 
the Fe XXV/XXVI best-fit temperatures obtained with CALDB 3.4 and CALDB 4.2 are nearly identical 
(see Fig.  \ref{acis_old_new_6070.fig}). This result supports the above suggestion that the Fe XXV/XXVI line ratio 
temperature is rather insensitive to smooth changes in the calibration of the effective area. 

\subsection{Accuracy of the energy resolution and redistribution calibration}
We examined how robust our Fe XXV/XXVI temperature measurements are when considering possible uncertainties in modelling 
of the time-dependent energy resolution and redistribution of the pn and MOS (Guainazzi, 2010). Following the work of 
Molendi et al. (2009), we added a multiplicative ``gsmooth'' component to the model, which convolves the input model with
a Gaussian kernel. We first fitted each cluster individually. This yielded rather large statistical uncertainties for the
width of the Gaussian kernel ($\equiv \sigma$) within which the clusters are rather consistent. On average, the value for 
$\sigma$ is 26 eV, and the rms scatter is 20 eV. Thus, within the scatter, these results are roughly consistent with the 
Gaussian width of 4e V reported by Molendi et al. (2009) for Perseus. 

However, the cluster observations of our sample were performed in a short period, early in the \xmm 
mission (during years 2000-2002) when the energy resolution was stable (e.g. Guainazzi 2010). We thus experimented 
by simultaneously fitting the spectra of all clusters including the gsmooth-component, while forcing $\sigma$ to be equal
in all data sets. For $\chi^2$ - comparison, we also repeated the fit without the gsmooth-component.
We found that the addition of the gsmooth-component improved the fit significantly: $\chi^{2}$ decreased from 297.0 to 
285.9 while the number of free parameters increased from 18 to 19 while using 345 spectral bins. The fit obtains a value
of 35$\pm$7 eV for the Gaussian width, i.e. consistent with that obtained when fitting each cluster individually above.
This value is consistent with measurements of the on-board pn calibration source (see, e.g., Guainazzi et al. 2010).
Thus, our results indicate a systematic underestimation of the pn resolution in the response matrix used in this paper. 
A similar exercise to our MOS data yielded a Gaussian width consistent with zero,  i.e. this effect is not apparent in 
MOS (consistent with Guainazzi et al., 2010). Thus this is not a physical broadening effect of the lines, but rather an 
instrumental problem of the pn. 

The pn temperatures decreased systematically with the inclusion of the gsmooth-component, by 2\% on average. This is 
slightly smaller than the value obtained when fitting each cluster individually. 
Thus we conclude that the effect of the uncertainties in the calibration of the energy resolution and redistribution 
at $\sim$ 6--7 keV render the measured Fe XXV/XXVI pn temperatures too high by $\sim$2\%.

\subsection{Calibration of the energy scale}
We then examined how possible problems in the calibration of the energy scale in the $\sim$6--7 keV band
impact the Fe XXV/XXVI temperature measurements. There are two principal factors affecting the energy reconstruction
of an event, namely Charge Transfer Inefficiency (CTI) and gain. CTI describes the transfer of charge as it is 
transported through the CCD to the output amplifiers during read-out. Gain is the conversion of the charge signal 
deposited by a detected photon from charge into energy (Guainazzi 2010). Note that possible uncertainties in the redshift 
measurement using the spectra of the galaxies of a given cluster (NASA Extragalactic Database) may contribute to the 
uncertainty of the energy scale because we keep the redshift fixed in the fits.

In order to examine the above effects, we used the ``gain fit'' option in XSPEC 11.3.2.ag package to modify the 
definition of the channel energies in the response matrix with an offset that does not depend on the energy. We fitted each
cluster individually and found that in average the offset parameter obtained values of 5 eV and 
16 for the pn and MOS, respectively. Considering the rms scatter, these values are consistent with the nominal accuracies of 
energy reconstruction of 10 eV and 5 eV for the pn and MOS, respectively, in the full energy range (Guainazzi 2010).

The modification of the gain offset renders the temperatures systematically lower, by  1\% (3\%) for the pn (MOS) on average.
Thus, we conclude that the uncertainties in the calibration of the energy scale of the EPIC instruments combined with the 
possible uncertainties of the cluster redshift measurements render the measured Fe XXV/XXVI EPIC temperatures too high by
$\sim$1--3\%.

\subsection{Details of the emission model}
In addition to the MEKAL model we adopted for the emission modelling, there also exists a commonly used emission model 
APEC (Smith et al., 2001). Differences on the details of the modelling of electron transitions may affect the interpreted 
temperature for a given line ratio data. We examined this by fitting the $\sim$6--7 keV band pn data using either MEKAL 
or APEC model. We found that the temperatures derived with APEC model are systematically lower than those obtained 
with the MEKAL model, by 2\% on average.

We also experimented with the choise of the element abundance tables by replacing our adopted one(Grevesse \& Sauval, 
1998) with that of Anders \& Grevesse (1989) or Lodders (2003). The differences in the derived $\sim$6--7 keV band 
temperatures are smaller than 0.1\%.

\end{appendix}

\end{document}